\documentclass[twocolumn,showpacs,preprintnumbers,amsmath,amssymb,superscriptaddress]{revtex4}

\usepackage{graphicx,amsmath,slashbox}

\newcommand{\unit}[1]{\,{\rm #1}}
\newcommand{\sub}[1]{_{\rm #1}}
\newcommand{\subup}[1]{^{\rm #1}}
\newcommand{\e}[1]{{\rm e}^{#1}}

\newcommand{\vctr}[1]{{\bf {#1}}}
\newcommand{\hc}{{\rm h.\,c.}}
\newcommand{\ann}[1]{#1^{\phantom{\dagger}}}
\newcommand{\cre}[1]{#1^{\dagger}}

\begin{document}

\title{Spin electric effects in molecular antiferromagnets}

\author{Mircea Trif}
\affiliation{Department of Physics, University of Basel,
             Klingelbergstrasse 82, CH-4056 Basel, Switzerland}
\author{Filippo Troiani}
\affiliation{CNR-INFM National Research Center S3  c/o Dipartimento
di Fisica via G. Campi 213/A, 41100, Modena, Italy}
\author{Dimitrije Stepanenko}
\affiliation{Department of Physics, University of Basel,
             Klingelbergstrasse 82, CH-4056 Basel, Switzerland}
\author{Daniel Loss}
\affiliation{Department of Physics, University of Basel,
             Klingelbergstrasse 82, CH-4056 Basel, Switzerland}

\date{\today}

\begin{abstract}

Molecular nanomagnets show clear signatures of coherent behavior and
have a wide variety of effective low-energy spin Hamiltonians suitable
for encoding qubits and implementing spin-based quantum information
processing.  At the nanoscale, the preferred mechanism for control of
quantum systems is through application of electric fields, which are
strong, can be locally applied, and rapidly switched.  In this work,
we provide the theoretical tools for the search for single molecule
magnets suitable for electric control.  By group-theoretical symmetry
analysis we find that the spin-electric coupling in triangular
molecules is governed by the modification of the exchange interaction,
and is possible even in the absence of spin-orbit coupling.  In
pentagonal molecules the spin-electric coupling can exist only in the
presence of spin-orbit interaction.  This kind of coupling is allowed
for both $s=1/2$ and $s=3/2$ spins at the magnetic centers.  Within
the Hubbard model, we find a relation between the spin-electric
coupling and the properties of the chemical bonds in a molecule,
suggesting that the best candidates for strong spin-electric coupling
are molecules with nearly degenerate bond orbitals.  We also
investigate the possible experimental signatures of spin-electric
coupling in nuclear magnetic resonance and electron spin resonance
spectroscopy, as well as in the thermodynamic measurements of
magnetization, electric polarization, and specific heat of the
molecules.   

\end{abstract}


\maketitle

\section {Introduction}

The control of coherent quantum dynamics is a necessary prerequisite
for quantum information processing.  This kind of control is achieved
through coupling of the internal quantum degrees of freedom of a
suitable micro- or mesoscopic system to an external classical or
quantum field that can readily be manipulated on the characteristic
spatial and temporal scales of the quantum system.  

The molecular nanomagnets (MNs) \cite{GSV06,GS03} represent a class of
systems that show rich quantum behavior.   At low energies, the MNs
behave as  a large spin or a system of only few interacting spins.
The behavior of this spin system can be designed to some degree by
altering the chemical structure of the molecules, and ranges from a
single large spin with high anisotropy barrier, to small collections
of ferro- or antiferromagnetically coupled spins with various
geometries and magnetic anisotropies.  This versatility of available
effective spin systems makes the MNs promising carriers of quantum
information  \cite{LL01}.  While the interaction with magnetic fields
provides a straightforward access to the spins in an MN, it is
preferable to use electric fields for the quantum control of spins,
since the electric fields are easier to control on the required short
spatial and temporal scales.  In this work, we explore the mechanisms
of spin-electric coupling and study the ways in which an MN with
strong spin-electric coupling can be identified.  

Quantum behavior of MNs is clearly manifested in the quantum tunneling
of magnetization
\cite{CG88,ASG+92,SGC+93,TLB+96,FST+96,WBH+97,TZB+97,BKR+03}.   A
prototypical example of quantum tunneling of magnetization is the
hysteresis loop of an MN with a large spin and high anisotropy
barrier.  The height of the barrier separating the degenerate states
of different magnetization leads to long-lived spin configurations
with nonzero magnetic moment in the absence of external fields.  The
transitions between magnetization states in the MN driven through a
hysteresis loop occur in tunneling events that involve coherent change
of a many-spin state.  These transitions have been observed as
step-wise changes in magnetization in single-molecule ferromagnets
\cite{GCP+94,TLB+96,FST+96,SOP+97,GSC00}.  Similar tunneling between
spin configurations are predicted in antiferromagnetic molecules
\cite{CL98,ML01}, and the observed hysteresis was explained in terms
of the photon bottleneck and Landau-Zener
transitions\cite{LL00,CWM+00,CWM+00b,WKS+02}.  The transitions between
spin states are coherent processes and show the signatures of
interference between transition paths \cite{LDG92,LL00b,LML03}, as
well as the effects of Berry phase in tunneling
\cite{LDG92,WS99,LL01b,LML03,GL07,GLM08}.  

Spin systems within molecular nanomagnets offer a number of attractive
features for studying the quantum coherence and for the applications
in quantum information processing \cite{LL01}.  A wide variety of spin
states and couplings between them allows for encoding qubits.
Chemical manipulation offers a way to modify the structure of
low-energy spin states \cite{TGA+05}.  Coherence times of up to $\sim
3\unit{\mu s}$ \cite{ARM+07} which can persist up to relatively high
temperatures of the order of few Kelvin are sensitive to the isotopic
composition of the molecule.  A universal set of quantum gates can be
applied in a system of coupled antiferromagnetic ring molecules,
without the need for local manipulation \cite{CSA+07}.  The presence
of many magnetic centers with the coupled spins allows for the
construction of spin cluster qubits that can be manipulated by
relatively simple means \cite{MLL03}.  In polyoxometalates, the spin
structure of the molecule is sensitive to the addition of charge, and
controlled delivery and removal of charges via an STM tip can produce
useful quantum gates \cite{LGE+07}.  Chemical bonds between the
molecules can be engineered to produce the permanent coupling between
the molecular spins and allow for interaction between the qubits
\cite{TCT+08}. 

Sensitivity of molecular state to the addition of charge was
demonstrated in the tunneling through single molecules \cite{HDF+06},
and used to control the spin state of a MN \cite{OMV+10}.  Transport
studies of the MNs can provide a sensitive probe of their spin
structure \cite{RWS06,RWH+06,LM06,GLM08,MCP+10}.

The most straightforward and traditional way of controlling  magnetic
molecules is by applying an external magnetic field.  With carefully
crafted ESR pulses, it is possible to perform the Grover algorithm, or
use the low-energy sector of the molecular nanomagnet as a dense
classical memory \cite{LL01}.  Unfortunately the approaches based on
magnetic fields face a significant drawback in the large-scale quantum
control application.  Typically, the quantum manipulation has to be
performed on the very short spatial and temporal scales, while the
local application of rapidly varying magnetic field presents a
challenging experimental problem.  For that reason, the schemes for
quantum computing tend to rely on modifying the spin dynamics that is
caused by intramolecular interaction, rather than on the direct
manipulation of spins \cite{TAC+05}.  

For the applications that require quantum control, the electric fields
offer an attractive alternative for spin manipulation in the molecular
nanomagnets \cite{TTS+08}.  One major advantage is that they can be
applied to a very small volume via an STM tip \cite{HLH06,BFB+08}, and
rapidly turned on and off by applying voltage pulses to the electrodes
placed close to the molecules that are being manipulated.  Switchable
coupling between different nanomagnets is essential for qubit
implementation.  At present, this can be implemented only locally, and
the interaction is practically untunable.  The use of microwave
cavities can offer a solution to this problem.  By placing the
nanomagnets inside a microwave cavity, one can obtain a fully
controllable, long-range interaction between them\cite{TTS+08}. This
coupling relies on the presence of a quantum electric  field inside
such a cavity, which mediates the interaction between distant
nanomagnets.  The interaction can be tuned by tuning each molecule in-
or out-of-resonance with the cavity field using local electric or
magnetic fields\cite{TTS+08}.  The spins, however, do not couple
directly to the electric fields, classical or quantum, and therefore
any electric spin manipulation is indirect, and involves the
modification of molecular orbitals or the spin-orbit interaction.  

The  description of the molecular nanomagnets  in terms of spins is an
effective low-energy theory that does not carry information about the
orbital states.  However, it is still possible to predict the form of
spin-electric coupling from symmetry considerations and single out the
molecules in which such a coupling is possible.  In particular, the
molecules with the triangular arrangement of antiferromagnetically
coupled spin-$1/2$ magnetic centers interact with external electric
field through chirality of their spin structure \cite{TTS+08,PPZ09}.
The same coupling of chirality to the external electric field was
derived for the triangular Mott insulators \cite{BBM+08}.

While the symmetry of a molecule sets the form of spin-electric
coupling, no symmetry analysis can predict the size of the
corresponding coupling constant.  The coupling strength will depend on
the underlying mechanism that correlates the spin and orbital states,
and on the detailed structure of low-energy molecular orbitals.  To
identify molecules that can be efficiently manipulated by electric
fields, it is necessary to perform an extensive search among the
molecules with the right symmetries and look for the ones that also
have a large coupling constant.  Unfortunately, this search has to
proceed by ab-initio calculations of the coupling constants for a
class of molecules of a given symmetry, or by an indiscriminate
experimental scanning of all of the available molecules.

In this paper, we contribute to the search for molecules that exhibit
strong spin-electric coupling.  Based on the symmetry analysis, we
identify the parameters of the spin Hamiltonian that can change in the
magnetic field, and cause spin-electric coupling.  We study the
mechanisms that lead to this coupling and describe the experiments
that can detect it.  

We will consider the spin electric coupling in the language of
effective model, namely either the spin Hamiltonian, or the Hubbard
model.  In reality the mechanism behind the spin-electric coupling
involves either the modification of the electronic orbitals in an
external field and the Coulomb repulsion of electrons, or the much
weaker direct spin-orbit coupling to the external fields.  A
derivation of spin-electric coupling from this realistic picture would
require the knowledge of electronic orbitals from an ab-initio
calculation, and the distribution of electric field within the
molecule.  Both of these problems require substantial computational
power, and can not be performed routinely.   Since the electric field
acts primarily on the orbital degrees of freedom, and the spin
Hamiltonian carries no information about the orbital states, we
provide a description in terms of a Hubbard model that still contains
some information about the orbital states.  We can then described the
properties of the molecule that allow for strong spin-electric
coupling in the language of orbitals that offers some intuitive
understanding of the underlying mechanisms of interaction.

We identify the response of an MN with spin-electric coupling in the
standard measurements of  ESR, nuclear magnetic resonance (NMR),
magnetization, polarization, linear magnetoelectric effect, and
specific heat measurements.

In Sec. II  we present a symmetry analysis of the spin-electric
coupling in the ring-shaped molecules with antiferromagnetic coupling
of spins.  In Sec. IV, we describe the MNs using the Hubbard model,
and relate the symmetry-based conclusions to the structure of
molecular orbitals. In Sec V, we analyze the experimental signatures
of spin-electric coupling, and present our conclusions in Sec. VI.

\section{Symmetry analysis of antiferromagnetic spin rings}

\begin{figure}[t]
\begin{center}
\includegraphics[width=7.5cm]{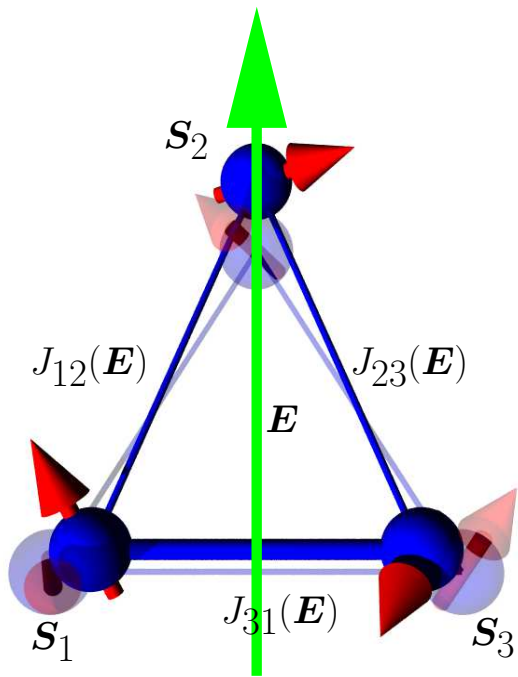}
\caption{(Color online) Schematics of the $s_i=1/2$ triangular
  molecule in electric field. The antiferromagnetic exchange couplings,
  represented by the bonds with thickness proportional to $J_{ii+1}$, are
  modified in electric field.  In the absence of electric field,
  exchange couplings are equal $J_{ii+1}=J_{jj+1}$, fade colors (grey
  online).  The full color (blue online) triangle represents the
  exchange interaction strengths in electric field.}
\label{triangle_field}
\end{center}
\end{figure}

Spin chains whose ground state multiplet consists of
two quasi-degenerate S = 1/2 doublets represent suit-
able candidates for the manipulation of the spin state
by pulsed electric fields. Such a ground-state multiplet
characterizes a number of frustrated spin rings, consisting
of an odd number of half-integer spins. In the following
we consider prototypical examples of such systems.

\subsection{Triangle of $s=1/2$ spins}

The low-energy properties of most molecular nanomagnets (MNs) are well
described in terms of spin degrees of freedom alone. Within the
spin-Hamiltonian approach, the coupling of external electric fields to
the molecule can be accounted by suitably renormalizing the physical
parameters.  In the following, we use the symmetry of the molecules to
calculate the changes of spin-Hamiltonian parameters, to identify the
system's eigenstates, and to deduce the allowed transitions.
Quantitative estimates of the parameters entering the spin Hamiltonian
require the use of ab-initio calculations \cite{BOM06}, or the
comparison with experiments.  The simplest example of a spin system
which may couple to an external electric field in a non-trivial way is
a triangle of s = 1/2 spins, like, for example, the Cu$_3$
MN\cite{CMN+06}.  The schematics of such a spin system in the presence
of an electric field is showed in Fig. \ref{triangle_field}.    Its
spin Hamiltonian, for the moment in the absence of any external fields
(magnetic or electric), reads:

\begin{equation}
H\sub{spin}=\sum_{i=1}^{N} J_{ii+1} \vctr{s}_i \cdot \vctr{s}_{i+1} +
\sum_{i=1}^N \vctr{D}_{ii+1} \cdot ( \vctr{s}_i \times \vctr{s}_{i+1}
),
\label{spin_hamiltonian}
\end{equation}
with $N=3$ and $\vctr{s}_4 \equiv \vctr{s}_1$ in the summation over
$i$. The first term  in Eq. (\ref{spin_hamiltonian}) represents the
isotropic Heisenberg exchange Hamiltonian with the exchange couplings
$J_{ii+1}$ between the spins $\vctr{s}_i$ and $\vctr{s}_{i+1}$, and
the second term represents the Dzyalozhinsky-Moriya (DM) interaction
due to the presence of spin-orbit interaction (SOI) in the molecule,
with the DM vectors $\vctr{D}_{ii+1}$.  The states of the spin $S=1/2$
triangle can be found by forming the direct product of the $SU(2)$
representations of three spins $S=1/2$:
$D_{tot}=D^{(1/2)\otimes3}=2D^{(1/2)}\oplus D^{(3/2)}$, meaning there
are eight states in total. The point group symmetry of the molecule is
D$_{3h}$\cite{CMN+06}, i.e. the triangle is assumed to be
equilateral. The D$_{3h}$ symmetry imposes the following restrictions
on the spin Hamiltonian parameters: $J_{ii+1}\equiv J$ and
$D^{x,y}_{ii+1}\equiv0$,  and $D^z_{ii+1}\equiv D_z$. However, if
lower symmetry is considered these restrictions will be relaxed. The
spin states in a form adapted to the rotational symmetry C$_{3}$ of
the system are
\begin{eqnarray}
|\psi_{M=1/2}^{(k)}\rangle&=&\frac{1}{\sqrt{3}}\sum_{j=0}^2\epsilon_{j}^kC_3^j\left|\downarrow\uparrow\uparrow\right\rangle\\
|\psi_{M=3/2}\rangle&=&\left|\uparrow\uparrow\uparrow\right\rangle, 
\label{spin_states}
\end{eqnarray} 
where $\epsilon_{j}=\exp{(2i\pi/3j)}$ and $j=0,1,2$. The states with
opposite spin projection $M'=-M$, i.e. with all spins flipped can be
written in an identical way (not shown).  These states are
already the symmetry adapted basis functions of the point group
D$_{3h}$. Moreover, these are eigenstates of the chirality operator 
\begin{equation}
C_z=\frac{1}{4\sqrt{3}}\vctr{s}_1\cdot(\vctr{s}_2\times\vctr{s}_3),
\label{chirality}
\end{equation}
with
$C_z|\psi^{(1,2)}_{M=\pm1/2}\rangle=\pm|\psi^{(1,2)}_{M=\pm1/2}\rangle$,
$C_z|\psi^{(0)}_{M=\pm1/2}\rangle=0$ and
$C_z|\psi_{M=\pm3/2}\rangle=0$. The above states in
Eq. (\ref{spin_states}) carry different total spin.  There are two
spin $S=1/2$ states, corresponding to $k=1,2$, and a spin $S=3/2$
state corresponding to $k=0$. Obviously, the states
$|\psi_{M=\pm3/2}\rangle$ have $S=3/2$. 

In an even-spin system, double valued point groups, instead of single
valued groups, are usually used in order to describe the states, the
splittings and the allowed transitions (magnetic or
electric)\cite{T06}.  In the presence of of spin-orbit interaction the
splittings can be accounted for either by single group analysis
(perturbatively), or by double group analysis (exact).  In the
following, we analyze the spectrum and the allowed transitions by both
single valued point group analysis and double valued point group
analysis.

\subsubsection{Single valued group analysis of the $s=1/2$ spin triangle}%

In the single valued point group D$_{3h}$, the states
$|\psi_{M=\pm1/2}^{(k)}\rangle$  with $k=1,2$ form the basis of the
two dimensional irreducible representation $E^{'}$, while the states
$|\psi_{M=\pm1/2}^{(0)}\rangle$, and the $|\psi_{M=\pm3/2}\rangle$
transform as $A_2^{'}$.  The allowed  electric transitions in the
system are determined by the transformation properties of the basis
states. 

The simplest and possibly the dominant  dependence of the spin
Hamiltonian on the applied electric field comes via the modification
of the exchange interactions, like depicted  in Fig. \ref{triangle_field}. This gives rise to the following term in
the spin Hamiltonian
\begin{equation}
\delta H_{0}(\vctr{E})=\sum_{i=1}^3\delta
J_{ii+1}(\vctr{E})\,\vctr{s}_i\cdot\vctr{s}_{i+1},
\label{E_modif_exchange}
\end{equation}
where $\delta J_{ii+1}(\vctr{E})\approx \vctr{d}_{ii+1}\cdot\vctr{E}$,
with $\vctr{d}_{ii+1}$ being vectors that describe the electric-dipole
coupling of the bond $\vctr{s}_i-\vctr{s}_{i+1}$ to the electric field
$\vctr{E}$ in leading order. There are three such vector parameters and
thus nine scalar parameters in total. However, symmetry will allow to
drastically reduce the number of free parameters by providing
relations between them.  The $S=3/2$ states of the unperturbed spin
Hamiltonian form the multiplet ${}^4A_2^{'}$, while the $S=1/2$ states
form two multiplets ${}^2E^{'}$. The electric dipole Hamiltonian is
$H\sub{e-d}=-e\sum_{i}\vctr{E}\cdot\vctr{r}_i\equiv
-e\vctr{E}\cdot\vctr{R}$, with $e$ standing for the electron charge,
$\vctr{r}_i$ being the coordinates of the $i$-th electron and
$\vctr{R}=\sum_i\vctr{r}_i$. The non-zero electric dipole matrix
elements of $H\sub{e-d}$ in the D$_{3h}$ symmetric molecule are 
\begin{equation}
\langle \psi^{(1,2)}_{M}|-ex|\psi^{(2,1)}_{M'}\rangle=i\langle
\psi^{(1,2)}_M|-ey|\psi^{(2,1)}_{M'}\rangle\equiv d\delta_{MM'},
\label{eld_dipole_matelem}
\end{equation} 
proportional to the effective electric dipole parameter $d$.  The
value of $d$ is not determined by symmetry, and has to be found by
some other means (ab-initio, Hubbard modeling, experiments, etc).  We
mention that all the other matrix elements are zero, e.g. $\langle
\psi^{(1,2)}_{M}|-ex|\psi^{(1,2)}_{M'}\rangle=i\langle
\psi^{(1,2)}_M|-ey|\psi^{(1,2)}_{M'}\rangle=0$, etc. We see that the
electric field acts only in the low-energy sector, which  allows us to
write the effective spin-electric  coupling Hamiltonian acting in the
lowest quadruplet as
\begin{equation}
H\sub{e-d}\subup{eff}=d\vctr{E}'\cdot\vctr{C}_{\parallel},
\label{effective_edipole}
\end{equation} 
where $\vctr{E}'=\mathcal{R}_{z}(7\pi/6-2\theta)\vctr{E}$,  with
$\mathcal{R}_{z}(\phi)$ describing the rotation with an angle $\phi$
about the $z$ axis, and $\theta$ is the angle between in-plane
component $\vctr{E}_{\parallel}$ of the electric field $\vctr{E}$ and
the bond $s_1-s_2$.  For  $\vctr{C}_{\parallel}=(C_x,C_y,0)$ we have
\begin{eqnarray}
C_x&=&\sum_{M}\left(|\psi^{(1)}_M\rangle\langle\psi^{(2)}_M|+|\psi^{(2)}_M\rangle\langle\psi^{(1)}_M|\right),\\ 
C_y&=&i\sum_{M}\left(|\psi^{(1)}_M\rangle\langle\psi^{(2)}_M|-|\psi^{(2)}_M\rangle\langle\psi^{(1)}_M|\right).\end{eqnarray}
The low-energy spectrum in the presence of electric field and the
related states can be expressed in terms of the spin Hamiltonian
Eq. (\ref{E_modif_exchange}), so that we find anisotropic variations
of the exchange coupling constants: 
\begin{equation}
\delta
J_{ii+1}(\vctr{E})=\frac{4d}{3}|\vctr{E}_{\parallel}|\cos{\left(\frac{2\pi}{3}i+\theta\right)},
\end{equation}
which depend on the angle $\theta$ and the projection of the electric
field $\vctr{E}$ on the plane of the triangle. In the $s_i=1/2$
triangle the $\vctr{C}$-operators can be written as 
\begin{eqnarray}
C_{x}&=&-\frac{2}{3}(\vctr{s}_1\cdot\vctr{s}_2-2\vctr{s}_2\cdot\vctr{s}_3+\vctr{s}_3\cdot\vctr{s}_1),\\   C_{y}&=&\frac{2}{\sqrt{3}}(\vctr{s}_1\cdot\vctr{s}_2-\vctr{s}_3\cdot\vctr{s}_1),
\label{parallel_chirality}
\end{eqnarray}
with $[C_i,C_j]=2i\epsilon_{ijk}C_k$ ($\epsilon_{ijk}$ are the
Levi-Civita symbols)\cite{TTS+08,BBM+08}. From the above relations we
can conclude that (i) only the electric field component perpendicular
to the bond and lying in the plane of the molecule gives rise to
spin-electric coupling and (ii) there is only one free parameter $d$
describing the coupling of the spin system to electric fields and
$\vctr{d}_{ii+1}=4d/3\left(\sin{(2i\pi/3)},\cos{(2i\pi/3)},0\right)$,
where $i=1,2,3$ labels the triangle  sites and $4\equiv1$. 
 
The SOI in a D$_{3h}$  symmetric MN is constrained by the
transformation properties of the localized orbitals. It reads
\begin{equation}
H\sub{SO}=\lambda\sub{SO}^{\parallel}T_{A_2}S_z+\lambda\sub{SO}^{\perp}(T_{E^{''}_+}S_-+T_{E^{''}_-}S_+),
\label{molecular_SOI}
\end{equation}
with $T_{\Gamma}$ being tensor operators transforming according to the
irreducible representation $\Gamma$\cite{T06}. The non-zero matrix
elements of this SOI Hamiltonian in the low-energy quadruplet read
$\langle
\psi^{(1,2)}_M|H\sub{SO}|\psi^{(1,2)}_{M'}\rangle=\pm
M\lambda\sub{SO}^{\parallel}\delta_{MM'}$ so that the SOI takes the
following effective form
\begin{equation}
H\sub{SO}=\Delta\sub{SO}C_zS_z,
\label{effective_SOI}
\end{equation}
with $\Delta\sub{SO}=\lambda\sub{SO}^{\parallel}$ and
$S_z=\sum_i^3s_i^z$. An effective  SOI Hamiltonian is obtained also
from the DM SOI Hamiltonian  in Eq. (\ref{spin_hamiltonian}). The
constraints $D^{x,y}_{ii+1}=0$ and $D^z_{ii+1}\equiv D_z$ on the DM
vectors due to  D$_{3h}$ symmetry of the molecule, give rise to the
same effective SOI in Eq. (\ref{effective_SOI}), with
$D_z=\lambda\sub{SO}^{\parallel}$. Thus, as expected, the molecular SOI
and the DM SOI give rise to the same effective SOI  Hamiltonian acting
in the low energy quadruplet. Like in the case of the electric dipole
parameter $d$, finding  $D_z (\lambda\sub{SO}^{\parallel})$ requires more
than symmetry, like ab-initio methods or experiments. The transverse
SOI, with interaction strength $\lambda\sub{SO}^{\perp}$ does not act
within the low-energy space, and its effect will appear only in higher
orders of perturbation theory in $1/J$.

\begin{figure}[t]
\begin{center}
\includegraphics[width=8.5cm]{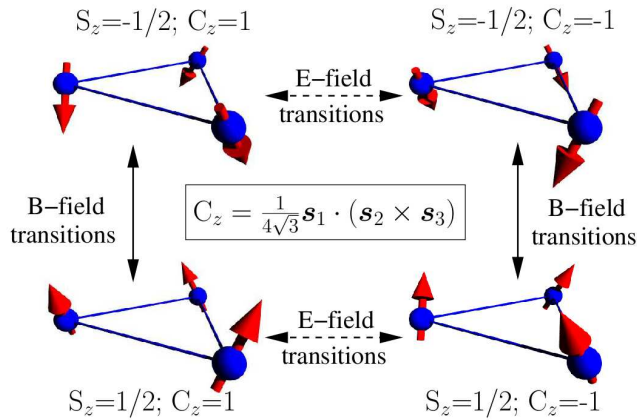}
\caption{The spin transitions in the $s_i=1/2$ triangle induced by
  electric and magnetic fields.  The electric field causes transitions
  between the states of opposite chiralities $C_z$ and equal spin
  projections $S_z$ (horizontal arrows), while the magnetic field
  instead causes transitions between the states of opposite spin
  projections $S_z$ and equal chiralities $C_z$ (vertical arrows).}
\label{chirality_trans}
\end{center}
\end{figure}

An external magnetic field couples to the spin via the Zeeman term
$H_Z=\vctr{B}\cdot\bar{\bar g}\vctr{S}$, with $\bar{\bar
  g}={\rm diag}\{g_{\parallel},g_{\parallel},g_{\perp}\}$ being the
$g$-factor tensor in D$_{3h}$. The full effective Hamiltonian
describing the low-energy quadruplet in the presence of SOI, electric
field and magnetic field  read
\begin{equation}
H\sub{eff}=\Delta\sub{SO} C_z S_z + \vctr{B} \cdot \bar{\bar g}
\vctr{S} + d \vctr{E}' \cdot \vctr{C}_{\parallel}.
\label{full_effective}
\end{equation}
Note that $[\vctr{C},\vctr{S}]=0$, and chirality and spin act as
independent spin $1/2$ degrees of freedom. Furthermore,  in the
absence of SOI the chirality $\vctr{C}$ and the spin $\vctr{S}$ evolve
independently. However, the SOI couples the two and provides with
means for electric control of both spin and chirality. Vice-versa,
magnetic fields can also couple to chirality due to SOI. Also, while
magnetic fields (time-dependent) cause transitions between states of
opposite spin projection $M$ but with the same chirality $C_z$, the
electric field does the opposite: it causes transitions between states
of opposite chirality $C_z$, but carrying the same $M$.    Full
control of the lowest quadruplet is thus realized in the presence of
both electric and magnetic fields, as can be seen in
Fig. \ref{chirality_trans}. 

%

\subsubsection{Double valued group states of the $s=1/2$ spin triangle}%

The double group representations allow to non-perturbatively describe
the magnetic and electric transitions in the presence of spin-orbit
interaction. The lowest quadruplet consists of two Kramers doublets,
one of them transforming like
$\bar{E}^{'}\sim(|-1/2\rangle,|1/2\rangle)$, and the other one
according to $\bar{E}^{''}\sim(|-3/2\rangle,|3/2\rangle)$. Here
$(|M\rangle,|-M\rangle)$  represent pairs  of eigenstates of a given
angular momentum $J\geq M$, with spin projection $\pm M$. For example,
if $M=1/2$, then $J=1/2,3/2,\dots$. The higher energy  states instead
($S=3/2$ states), transform now not as $A_2^{'}$, but as $\bar{E}^{'}$
($M=\pm1/2$) and as $\bar{E}^{''}$ ($M=\pm3/2$).  Thus, the $S=1/2$
states mix with the $S=3/2$ states, but only the ones transforming
according to the same representations, i.e. there is no mixing between
$\bar{E}^{'}$ and $\bar{E}^{''}$ due to spin-orbit interaction. The
magnetic dipole transitions take place between $\bar{E}^{'}$ and
$\bar{E}^{''}$, and within  $\bar{E}^{'}$ and $\bar{E}^{''}$,
respectively, while electric dipole transitions take place only
between $\bar{E}^{'}$ and $\bar{E}^{''}$. The selection rules for the
electric transitions are $\Delta M=\pm2$, while for the magnetic
transitions these are $\Delta M=0\pm1$.  We see that within the double
group analysis, i.e. in the presence of SOI,  there are allowed
electric dipole transitions also within the $S=3/2$ subspace.

Using both the single group and double group analysis we can pinpoint
to the transitions that arise in the absence or only in the presence
of SOI. Therefore,  the electric dipole transitions present in the
single-group are a consequence of the modified exchange interaction,
and can arise even in the absence of SOI, while the ones that show up
only in the double group analysis are a consequence of the SOI (or
modification of SOI in electric field).  

We now can  establish several selection rules for the SOI, electric
field and magnetic field induced transitions. Note that the above
analysis was exact in SOI. However, it instructive to treat electric
field,  magnetic fields and SOI on the same footing. First, we find
that the electric dipole transitions fulfill the selection rules
$\Delta C_z=\pm1$ and $\Delta S_z=0$, meaning that  electric field
only couples  states within the lowest quadruplet. The SOI transitions
show a richer structure. We  can separate the SOI interaction in two
parts: the perpendicular SOI, quantified by $D_z$ in the DM
interaction Hamiltonian, and the in-plane SOI, quantified by $D_{x,y}$
in the DM interaction Hamiltonian, respectively. By doing so, we find
that the  $D_z$ SOI terms obey the selections  rules  $\Delta C_z=0$
and $\Delta S_z=0$, while for the $D_{x,y}$ terms we get the selection
rules $\Delta C_z=\pm1$ and $\Delta S_z=\pm1$.  We see in-plane SOI
($D_{x,y}$ terms) do not cause any splitting in the ground state and
can lead to observable effects only in second order in perturbation
theory in $D_{x,y}/J$. Also, note that if $\sigma_h$ symmetry is
present, $D_{x,y}\equiv0$ and thus there are no in-plane SOI effects
at all.   Modification of these terms due to an in-plane external
electric field $\vctr{E}$, however,   lead to different selection
rules: changes of $D_z$ terms lead to $\Delta C_z=\pm1$ and $\Delta
S_z=0$, while modification of $D_{x,y}$ lead to $\Delta C_z=0,\pm2$
and $\Delta S_z=\pm1$.  The magnetic field transitions obey the
selection rules $\Delta S_z=0,\pm1$ and $\Delta C_z=0$.  Thus, we can
make clear distinction between pure electric field transitions,
SOI-mediated electric transitions and magnetic transitions.  This
distinction between the electric and magnetic field induced
transitions could be used   to extract the spin-electric coupling
strength parameter $d$ from spectroscopic measurements.

\subsection{Spin $s=3/2$ triangle}%

The spin $s=3/2$ triangle has a  more  complex level structure than
the $s=1/2$ triangle due to its higher spin. The spin  Hamiltonian,
however,  is similar to the  one in Eq. (\ref{spin_hamiltonian}) for
$s=1/2$, and   the reduction of the representation of three spins
$S=3/2$  is $D_{tot}=D^{(3/2)\otimes3}=2D^{(1/2)}\oplus
4D^{(3/2)}\oplus 3D^{(5/2)}\oplus 2D^{(7/2)}\oplus D^{(9/2)}$, a
total of $64$ spin states. The total number of irreducible
representations is the same as in the $s=1/2$ case, and we need only
to identify these basis states in terms of the spin states.   The
$s=3/2$ triangle states can be defined according to their
transformation properties under three-fold rotations C$_{3}$ in
D$_{3h}$ and are of the following form
\begin{eqnarray}
|\psi_{M}^{(k,i)}\rangle&=&P_{k}^3|M,i\rangle,\\  P_k^3&=&\frac{1}{\sqrt{3}}\sum_{j=0}^2\epsilon_{j}^kC_3^j,
\end{eqnarray}
where $\epsilon_{j}^k=\exp{(2i\pi jk/3)}$, $C_3^j$ are the $3$-fold
rotation of order $j$, and $j,k=0,1,2$. The states
$|M,i\rangle\equiv|\sigma_1\sigma_2\sigma_3\rangle$ represent all
possible states ($i$ states in total) with a given spin projection
$M(\equiv\sum_{k}\sigma_k)$ that cannot be transformed into each other
by application of the rotation operator $C_3^j$. These states are
showed in Table \ref{spin_three_half_states}. 
\begin{table}[h]
\centering
\begin{tabular}{| c | c | c | c | c |}
\hline \backslashbox{M}{i} & 1 & 2 & 3 & 4 \\
\hline 1/2 & $\left|\downarrow\uparrow\uparrow\right\rangle$ &
$\left|\Uparrow\downarrow\downarrow\right\rangle$ &
$\left|\Downarrow\Uparrow\uparrow\right\rangle$ &
$\left|\Downarrow\uparrow\Uparrow\right\rangle$\\ 3/2 &
$\left|\Downarrow\Uparrow\Uparrow\right\rangle$ &
$\left|\downarrow\uparrow\Uparrow\right\rangle$ &
$\left|\downarrow\Uparrow\uparrow\right\rangle$ &
$\left|\uparrow\uparrow\uparrow\right\rangle$ \\ 5/2 &
$\left|\Uparrow\uparrow\uparrow\right\rangle$ &
$\left|\downarrow\Uparrow\Uparrow\right\rangle$ & 0 & 0\\ 7/2 &
$\left|\uparrow\Uparrow\Uparrow\right\rangle$ & 0 & 0 & 0\\ 9/2 &
$\left|\Uparrow\Uparrow\Uparrow\right\rangle$  & 0 & 0 & 0
\\ \hline\end{tabular}
\caption{Non-symmetry adapted states of the $s=3/2$ spin triangle.  We
  use $|\Uparrow(\Downarrow)\rangle=|\pm3/2\rangle$.}
\label{spin_three_half_states}
\end{table}

The corresponding states with all spins flipped, namely with
$M^{'}=-M$, can be written in a similar form (not shown).  Having
identified the symmetric states in terms of the spin states, we
proceed to analyze the allowed transitions induced in the spin systems
by magnetic and electric field, both  within the single valued group
and double valued group representations.

\subsubsection{Single valued group states of the $s=3/2$ triangle}%

The above states are basis of the point group D$_{3h}$, but not
eigenstates of the total spin operator $\vctr{S}^2$, i.e. they do not
have definite total spin. However, linear combinations of states of a
given total spin projection $M$ and a given 'chiral' numbers $k$
become eigenstates of $\vctr{S}^2$. The total spin eigenstates  can be
written as
$|\psi_{S,M}^{(k)}\rangle=\sum_{l(M)}a_{k,l}^S|\psi_{M}^{(k,l)}\rangle$,
where $l(M)$ is the number of different states with a given $M$. The
coefficients $a_{k,l}$ are to be identified so that these states
satisfy
$\vctr{S}^2|\psi_{S,M}^{(k)}\rangle=S(S+1)|\psi_{S,M}^{(k)}\rangle$, with
$S=1/2,3/2,5/2,7/2,9/2$. The states with $k=0$ are all transforming
according to the $A_2^{'}$ representation, while the states with
$k=1,2$ are organized in doublets, being the bases of the two
dimensional representation $E^{'}$. However, as mentioned above,
different combinations of symmetry adapted states carry different
total spin $S$. The magnetic and electric transitions are similar to
the ones in the $s=1/2$ triangle, in the absence of SOI. The electric
field causes transitions only between states with the same $M$ and
$S$, but opposite chirality
$C_z=\frac{1}{2\sqrt{3}}\vctr{s}_1\cdot(\vctr{s}_2\times\vctr{s}_3)$
(this is different from the triangle with $s_i=1/2$ spins in each of
the vertices). As for the $s=1/2$ spin triangle, there are electric
dipole transitions within the spin system even in the absence of
SOI. The ground states is four-fold degenerate consisting of two
$S=1/2$ eigenstates
\begin{eqnarray}
|\psi_{M=1/2}^{(1)}\rangle&=&\frac{1}{\sqrt{10}}\bigg(|\psi_{M=1/2}^{(1,1)}\rangle+\sqrt{3}|\psi_{M=1/2}^{(1,2)}\rangle\nonumber\label{three_half_gstates_1}\\ &-&(\epsilon_1-\epsilon_2)(|\psi_{M=1/2}^{(1,3)}\rangle-|\psi_{M=1/2}^{(1,4)}\rangle)\bigg),\\   \psi_{M=1/2}^{(2)}\rangle&=&\frac{1}{\sqrt{10}}\bigg(|\psi_{M=1/2}^{(2,1)}\rangle+\sqrt{3}|\psi_{M=1/2}^{(2,2)}\rangle\nonumber\\  &+&(\epsilon_1-\epsilon_2)(|\psi_{M=1/2}^{(2,3)}\rangle-|\psi_{M=1/2}^{(2,4)}\rangle)\bigg).
\label{three_half_gstates_2}
\end{eqnarray}
We see that, as opposed to the  $s=1/2$ triangle,   the lowest states
are given by linear combinations of the several $M=1/2$ symmetry
adapted  states (the $M=-1/2$ states are obtained by flipping the
spins in the states in Eqs. (\ref{three_half_gstates_1}),
(\ref{three_half_gstates_2}).  This, however, does not modify the
conclusions regarding the electric and magnetic transitions in the
absence of SOI, these being given by the same rules as in the $S=1/2$
triangle: electric-field induced transitions between the states of
opposite chirality $C_z$ and the same spin  projection $M$. The lowest
states are still organized as spin and chirality eigenstates that are
split  in the presence of SOI as in the previous case. 

In the original spin Hamiltonian in Eq. (\ref{spin_hamiltonian}) the
electric field causes modification of the spin Hamiltonian
parameters. As for the spin $s=1/2$ triangle, the strongest effect
comes from modification of the isotropic exchange interaction, so that
\begin{equation}
\delta H_0(\vctr{E})=\sum_{i=1}^3\delta
J_{ii+1}(\vctr{E})\vctr{s}_i\cdot\vctr{s}_{i+1},
\label{three_half_modif}
\end{equation} 
with $\delta J_{ii+1}(\vctr{E})=dE\cos{(2\pi i/3+\theta)}$, where
$\theta$ is the angle between the projection of the external electric
field $\vctr{E}$ to the molecule's plane and the
$\vctr{s}_1-\vctr{s}_2$ bond, and $i=0,1,2$. The effect of the
electric field on the lowest quadruplet is found to be similar to the
spin $s=1/2$ case. While the SOI splits the two chiral states without
mixing them (at least in lowest order), the electric field, on the
other hand, mixes the chiral states.  The effective  Hamiltonian
acting in the lowest quadruplet reads
\begin{equation}
H\sub{eff}=\Delta\sub{SO}C_zS_z+\vctr{B}\cdot\bar{\bar{g}}\vctr{S}+d'\vctr{E}\cdot\vctr{C}_{\parallel}.
\label{eff_Ham_three_half}
\end{equation}  
Above, $d'=3d/2$, $\vctr{C}_{\parallel}=(C_x,C_y,0)$, with
$C_x=\sum_M|\psi^{(1)}_M\rangle\langle\psi^{(2)}_M|+|\psi^{(2)}_M\rangle\langle\psi^{(1)}_M|$
and
$C_x=i\sum_M(|\psi^{(1)}_M\rangle\langle\psi^{(2)}_M|-|\psi^{(2)}_M\rangle\langle\psi^{(1)}_M|)$,
and $\Delta\sub{SO}$ stands for the SO splitting. However, in this
situation the in-plane chirality operators $C_{x,y}$ cannot be written
in a simple form as a function  of the individual spin operators, as
opposed to the $s=1/2$ triangle.

\subsubsection{Double valued group states of the $s=3/2$ triangle}%

The double group representation allows to identify the  couplings
between different spin states induced by the SOI and to identify  the
allowed magnetic dipole transitions.  Due to SOI, the  electric field
induced spin transitions will take place also outside the spin
quadruplet. In the absence of extra degeneracies (induced, for
example, by external magnetic fields), however,  these transitions are
strongly reduced due the gap of the order $J$. We can then  focus, as
for the $S=1/2$ triangle, only on the lowest quadruplet.  These states
are organized in two Kramer doublets of the form
$(|M\rangle,|-M\rangle)$, one transforming  as
$\bar{E}^{'}\sim(|1/2\rangle,|-1/2\rangle)$ and the other one as
$\bar{E}^{''}\sim(|-3/2\rangle,|3/2\rangle)$. Here again,
$(|M\rangle,|-M\rangle)$ represent angular momentum $J\leq M$
eigenstates with spin projection $\pm M$.

 As in the case of the $s=1/2$  triangle, the electric field induced
 transitions  take place between $\bar{E}^{'}$ and $\bar{E}^{''}$,
 with  the selection rules  $\Delta M=\pm2$. Magnetic transitions
 instead take  place both within and between $\bar{E}^{'}$ and
 $\bar{E}^{''}$, satisfying the selection rules  $\Delta M=0,\pm 1$. 

 If we now treat the SOI, electric field and magnetic fields on the
 same footing, we arrive at the same selection rules as for the
 $s=1/2$ triangle, namely  $\Delta C_z=\pm1$ and $\Delta S_z=0$ for
 electric transitions,  $\Delta C_z=0,\pm1$ and $\Delta S_z=0,\pm1$
 for SOI transitions, and $\Delta C_z=0$ and $\Delta S_z=0,\pm1$ for
 magnetic transitions, respectively.

\subsection{Spin $s=1/2$ pentagon}%

%
%
\begin{figure}[t]
\begin{center}
\includegraphics[width=8.5cm]{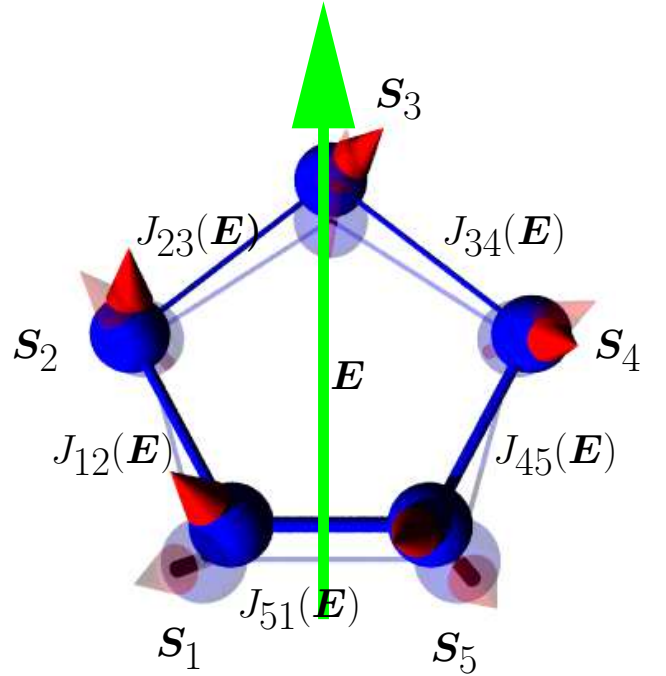}
\caption{(Color online) Schematics of a pentagonal spin ring molecule
  in electric field $\vctr{E}$, light (green) arrow.  The molecule in
  the absence of electric field is depicted in fade colors, while the
  full colors represent the molecules in electric field.  Thickness of
  the bonds represents the strength of antiferromagnetic exchange
  interaction between the spins.  An electric field modifies the
  strengths of spin exchange couplings $J_{ii+1}$.}
\label{pentagon_field}
\end{center}
\end{figure}

We now analyze  the spin-electric coupling in  a pentagonal  molecule
with a spin $s=1/2$ in each of the vertices, like depicted
schematically in Fig. \ref{pentagon_field}.  As in the case of the
spin triangle, an external electric field $\vctr{E}$ gives rise to
modification of exchange interaction $J_{ii+1}$ in
Eq. (\ref{spin_hamiltonian}).  However,  the net spin-electric
coupling in the lowest spin sector can only be mediated by SOI. i.e.
via the  DM interaction (which can be also modified in the presence of
the $\vctr{E}$-field).  

To make the analysis simpler, we assume in the following that the
pentagonal spin molecule possesses a $D_5$ point group symmetry, thus
no horizontal reflection plane $\sigma_h$. However, no generality is
lost, since lower symmetry implies more allowed transitions in the
spin system. If, for example, in the lower symmetric situation some
transitions are forbidden, these transitions will be forbidden in the
higher symmetry case.  The Hamiltonian is given in
Eq. (\ref{spin_hamiltonian}) with $N=5$.  The states of the pentagon
are found from the product of the individual spin representations
$D_{tot}=D^{(1/2)\otimes5}=5D^{(1/2)}\oplus 4D^{(3/2)}\oplus
D^{(5/2)}$,  meaning there are $32$ spin states in total. As before,
these states can be organized in a symmetry adapted basis in the
following way
\begin{eqnarray}
|\psi_{M}^{(k,i)}\rangle&=&P_5^k|M,i\rangle,\\ P_5^k&=&\frac{1}{\sqrt{5}}\sum_{j=0}^4\epsilon_{j}^kC_{5}^j,
\end{eqnarray}
where $\epsilon_{j}^k=\exp{(2i\pi jk/5)}$ with $k,j=0,\dots,5$, $C_5^j$
are the $5$-fold rotations of order $j$. The states
$|M,i\rangle\equiv|\sigma_1\sigma_2\sigma_3\sigma_4\sigma_5\rangle$
represent all possible states ($i$ states in total) with a given spin
projection $M(\equiv\sum_{k}\sigma_k)$ that cannot be transformed into
each other by application of the rotation operator $C_5^j$.  These
states are showed in in Table \ref{spin_pentagon_states}  
\begin{table}[h]
\centering
\begin{tabular}{| c | c | c | }
\hline \backslashbox{M}{i} & 1 & 2 \\
\hline 1/2 &
$\left|\uparrow\downarrow\uparrow\downarrow\uparrow\right\rangle$ &
$\left|\uparrow\downarrow\downarrow\uparrow\uparrow\right\rangle$
\\ 3/2 &
$\left|\downarrow\uparrow\uparrow\uparrow\uparrow\right\rangle$ & 0
\\ 5/2 & $\left|\uparrow\uparrow\uparrow\uparrow\uparrow\right\rangle$
& 0 \\ \hline
\end{tabular}
\caption{Spin $s=1/2$ pentagon non-symmetry adapted states.}
\label{spin_pentagon_states}
\end{table}
and the corresponding states with all spins flipped,
i.e. $M\rightarrow-M$ states (not shown). In the absence of SOI there
is no mixing of different $k$ states, i.e. the chirality is a good
quantum number. In this case the chirality is quantified by the
operator
$C_z=1/(2\sqrt{5+2\sqrt{5}})\sum_{i}\vctr{s}_i\cdot(\vctr{s}_{i+1}\times\vctr{s}_{i+2})$
(the prefactor is chosen for convenience; see below). As in the
$s=1/2,3/2$ spin triangles,  the above states are not yet the
eigenstates of the  Hamiltonian and we have to solve the equation
$S^2|\psi_{S}^{(i)}\rangle=S(S+1)|\psi_{S}^{(i)}\rangle$, with
$|\psi_{S}^{(i)}\rangle=\sum_{k(M)}a^S_{k,i}|\psi_{M}^{(k,i)}\rangle$. The
ground state is spanned, again, by four states, two Kramers doublets
with spin $S=1/2$. In the following we inspect the level  structure of
these four states in terms of the above symmetry adapted states.

\subsubsection{Single valued group $s=1/2$ pentagon}%

We focus here only on the four lowest energy states, which are two
pairs of $S=1/2$ states. The first (second) pair is given  by linear
combination of  states with chirality $k=1$ ($k=4$) and spin
projection $M=\pm1/2$. We obtain
\begin{eqnarray}
&|\psi_{S=1/2,M=\pm1/2}^{(k)}\rangle&=\frac{1}{\sqrt{3}}\bigg(\frac{1}{\displaystyle{2\cos{\left(\frac{2k\pi}{5}\right)}}}|\psi_{M=\pm1/2}^{(k,1)}\rangle\nonumber\\ &+&2\epsilon_2^k\cos{\left(\frac{2k\pi}{5}\right)}|\psi_{M=\pm1/2}^{(k,2)}\rangle\bigg),
\label{five_half_gstates}
\end{eqnarray}
so that
$C_z|\psi_{M=\pm1/2}^{(k)}\rangle=(-1)^k|\psi_{M=\pm1/2}^{(k)}\rangle$.
These states (for a given $M$ projection) form  the basis of the two
dimensional irreducible representation $E_1$. We are now in positions
to investigate the allowed electric dipole transitions within this
lowest subspace.  The in-plane electric dipole $\vctr{d}=(d_x,d_y)$
forms a basis of the irreducible representation $E_1$ in D$_{5}$. By
calculating the product  $E_1\otimes E_1\otimes E_1=2E_1\oplus2E_2$ we
see that the totally symmetric representation $A_1$ of D$_{5}$ is
absent. Therefore, there are {\it no} electric dipole transitions
within the four dimensional subspace in the absence of SOI.

As in the previous two cases, the coupling of the spin Hamiltonian to
electric field comes via modification of the spin Hamiltonian
parameters. If only the modification of the isotropic  exchange
Hamiltonian is taken into account, the spin-electric Hamiltonian takes
the same form as in Eq. (\ref{effective_edipole}), with $\delta
J_{ii+1}(\vctr{E})=dE\cos{(2i\pi/5+\theta)}$, $i=1\dots5$. The
parameter $d$ quantifies the electric dipole coupling of each of the
bonds and $\theta$ is the angle between the electric field $\vctr{E}$
and the bond $\vctr{s}_1-\vctr{s}_2$. Note that $d$ is in principle
non zero in D$_{5}$ point group symmetry. However, the matrix elements
of the spin-electric Hamiltonian within the lowest quadruplet are all
zero, i.e. $\langle\psi_{S=1/2,M}^{(k)}|\delta
H\sub{e-d}(\vctr{E})|\psi_{S=1/2,M'}^{(k')}\rangle\equiv0$. This means
that electric field has no effect on the lowest quadruplet, as found
out also by purely symmetry arguments. Therefore, we may expect that
the spin-electric coupling in pentagonal spin molecule is caused by SO
effects.

\subsubsection{Double valued group $s=1/2$ pentagon}%

Double valued group analysis allows identifying of the level structure
and the allowed transitions in the presence of SOI and magnetic
fields.  The lowest four states  in the double group $D_5^{'}$ are
described by the two dimensional irreducible representations
$\bar{E}_1^{'}\sim(|-1/2\rangle,|1/2\rangle)$ and
$\bar{E}_1^{''}\sim(|-3/2\rangle,|3/2\rangle)$, respectively. Since
both the magnetic  $\vctr{\mu}$ and electric $\vctr{d}$ dipoles
transform as $E_1$ in $D_5^{'}$, both electric and electric
transitions will take place between the same pair of states. The
products of the irreducible  representations that labels the states in
the low-energy quadruplet read:
$\bar{E}_1^{'}\otimes\bar{E}_2^{''}=E_{1}\oplus E_2$,
$\bar{E}_1^{'}\otimes\bar{E}_1^{'}=A_1\oplus A_2\oplus E_1$ and
$\bar{E}_2^{'}\otimes\bar{E}_2^{'}=A_1\oplus A_2\oplus E_2$. These
equalities  imply the same selection rules in the lowest subspace  as
for the spin triangle case: $\Delta M=\pm2$
($|\pm1/2\rangle\leftrightarrow|\mp3/2\rangle$) for electric dipole
transitions, and $\Delta M=\pm1$
($|\pm1/2\rangle\leftrightarrow|\mp1/2\rangle$ and
$|\pm1/2\rangle\leftrightarrow|\pm3/2\rangle$), for the magnetic ones.

The main  feature of pentagonal spin ring is the absence of electric
dipole transitions in the lowest quadruplet in the absence of
SOI. This is to be contrasted to the spin triangle case, where
spin-electric coupling exists in the ground state even in the absence
of SOI. This feature  finds its  explanation from the interplay
between  the selection rules for electric  field transitions and the
ones for the SOI.  In fact, these selection rules  are by no means
different from the triangular spin rings. Since the ground  state is
spanned by four states with chirality $C_z=1,4$ and spin $S_z=\pm1/2$,
we see that the condition $\Delta C_z=\pm1$ for the electric field
transitions implies no electric field coupling within the ground
state! In the presence of SOI though, spin electric coupling is still
possible, but it will be  $~(D_{x,y}/J)$ times smaller than in
triangles. Spin-electric coupling can arise also via modification of
the DM vectors $D_{x,y,z}$ in electric field. However, the selection
rules for this transitions are,  like for the triangle, $\Delta
C_z=0,\pm2$ and $\Delta S_z=0,\pm1$. This  means direct splitting in
the ground state, and thus we expect that for pentagon spin ring the
electric dipole response will be much weaker.

\section{Hubbard model of a molecular nanomagnet}

Spin-Hamiltonian models of molecular nanomagnets are based on
the assumption that the spins on magnetic centers are the only
relevant degrees of freedom. This assumption of fully quenched and
localized orbitals allows for the relatively simple predictions of
spin structure in the low-energy states of the molecule. However,
since the orbital dynamics plays a crucial role in spin-electric
coupling, spin-Hamiltonian models are unable to predict the
corresponding coupling constants. In this Section, we relax the 
assumption of
quenched and localized orbitals and treat the orbital degrees of
freedom of electrons on magnetic ions within a Hubbard model.  
This provides an intuitive picture of spin-electric coupling in 
terms of the deformation of the molecular orbitals induced by the
external field. 
Besides, in the limit of strong quenching of the 
orbitals, the Hubbard model reproduces a spin Hamiltonian, similar 
to the results found in the studies of cuprates 
\cite{M60,SEA92,YHE+94} and multiferroics 
\cite{SD06,DYY+09}.  
In particular, we find the relation between modifications of the 
electronic hopping matrix elements induced by the field and that
of the spin-electric coupling in the spin Hamiltonian,
thus providing a guide for the estimate of the size of spin-electric
coupling in a molecule.   

The outline of the present Section is the following.  
In Subsection \ref{sub:hubsym}, we introduce the Hubbard model of 
a spin chain with the shape of regular $n$-tangon, and derive the
resulting symmetry constraints for the hopping parameters. 
In Subsection \ref{sub:hcu3} we assume a direct electron hopping 
between magnetic sites, and derive the spin Hamiltonian of a 
spin triangle from the Hubbard model, in the limit of large on-site 
repulsions; we thus express the coupling to electric fields in terms of 
the Hubbard-model parameters.  
In Subsection \ref{sub:bridge}, we introduce a Hubbard model of a
magnetic coupling in the case where this is mediated by a non-magnetic 
bridge between the magnetic centers; also in this case, we find a 
connection between the modification of the bridge and spin-electric 
coupling.

\subsection{Parameters of the Hubbard model of molecular nanomagnets}
\label{sub:hubsym}

Magnetic properties of molecular nanomagnets are governed by the
spin state of few electrons in the highest partially occupied atomic
orbitals, split by the molecular field.  The spin
density is localized on the magnetic centers
\cite{PKP06}, and thus the low-energy magnetic
properties are correctly described by quantum models of interacting
localized spins \cite{B09,B09b}.  

The response of molecular nanomagnets to electric fields, as a matter
of principle, does not have to be governed by the electrons occupying
the same orbitals that determine the molecule's spin.  However, the
quantum control of single molecule magnets by electric fields depends
on the electrons that both react to electric fields and produce the
magnetic response.  Therefore, the models of molecular nanomagnets
that consider only few orbitals can provide useful information about
the electric control of spins.  

Hubbard model 
provides a simplified description of orbital degrees of freedom by
including only one or few localized orbitals on each magnetic center.
Furthermore, the interaction between electrons is accounted for only
by introducing the energies of the atomic configurations with
different occupation numbers.  The Hubbard model of the MN is given
by: 
\begin{equation}
\label{eq:hubbardMolecule}
\begin{split}
H\sub{H} =& \left[ \sum_{i,j}\sum_{\alpha,\beta} \cre{c}_{i\alpha}
  \left( t \delta_{\alpha \beta} + \frac{ i \vctr{P}_{ij}}{2} \cdot
       {\boldsymbol{\sigma}}_{\alpha \beta} \right) \ann{c}_{j\beta} +
       \hc \right] \\ &+ \sum_{j} U_j \left( n_{j\uparrow},
n_{j\downarrow} \right).
\end{split}
\end{equation}
where  $\cre{c}_{j \sigma}$ ($\ann{c}_{j \sigma}$) creates
(annihilates) an electron with spin $\sigma = \uparrow, \downarrow$ on
the orbital localized on $j$th atom, and  $n_{j\sigma} = \cre{c}_{j
  \sigma} \ann{c}_{j \sigma}$  is the corresponding number operator.
Model parameters $U_{j}$, describe the energy of
$n_{j\uparrow(\downarrow)}$ spin up(down) electrons electrons on the
site $j$.  Hopping parameters $t_{i j}$, $\vctr{P}_{i j}$ describe the
spin-independent and spin-dependent hopping between sites $i$ and $j$.

We assume that the largest energy scale is the splitting between the 
energy of the highest occupied atomic orbital and lowest unoccupied one, 
induced by the molecular crystal field: this justifies the inclusion 
of one orbital only for each magnetic center. The on-site repulsion
energy is the next largest energy scale in the problem, being $U_j$ 
larger than the hopping coefficients. Amongst these, processes 
involving states of different spin, mediated by spin-orbit interaction, 
are described by the $x$ and $y$ components of $\vctr{P}_{ij}$. 
The parameters $P_{ij;z}$, instead, describe the difference of the 
hopping matrix elements between spin-up and spin-down electrons. 
In the following, we shall consider both the case where electron 
hopping takes place directly between neighboring magnetic ions and 
that where the magnetic interaction is mediated by bridges of 
non-magnetic atoms. The Hubbard Hamiltonian can be approximated by a 
spin Hamiltonian model in the limit $|t_{i j}|, |\vctr{P}_{i j}| \ll U_j$.
The symmetry constraints on the spin Hamiltonian parameters can be 
deduced from those on the Hubbard model parameters \cite{M60}.
If the spin-independent hopping dominates ($|t| \gg |\vctr{P}|$), the
resulting spin Hamiltonian will contain the Heisenberg exchange terms 
and a small additional spin-anisotropic interaction.  
If $|t| \gtrsim |\vctr{P}|$, the size
of spin-dependent interactions in the spin Hamiltonian will be
comparable to the Heisenberg terms.  Both these cases appear in the
molecule nanomagnets \cite{CMN+06,CWM+00,CCG+05b,LBH+08}. 

Symmetry of the molecule imposes constraints to the Hubbard model,
thus reducing the number of free parameters. The on-site repulsion
parameters $U_j$ are equal for all equivalent magnetic ions.  In the
molecules of the form of regular $n$-tagon, all of the
spin-independent hopping parameters are equal, due to the $C_n$
symmetry. The spin-dependent hopping elements are related by both the
full symmetry of the molecule and the local symmetry of localized
orbitals.  For example, in the case of localized orbitals  in a
regular polygon that are invariant under the local symmetry group of
the magnetic center,
\begin{equation}
\label{eq:Pxcn}
P_{j,j+1;x} = \exp\left[ i\frac{2\pi(j-k)}{n} \right] P_{k,k+1;x},
\end{equation}
with the convention that site $n+1$ coincides with site $1$.  In this
case, there is only one free parameter that determines all of the
$P_{x}$ matrix elements.  Therefore, the regular $n-$tagon molecule in
the absence of external electric and magnetic fields can be described
by a Hubbard model, with five independent parameters: $U$, $t$,
$\vctr{P}_{12}$. In addition,  the $\sigma\sub{v}$ symmetry, if
present will impose $\vctr{P}_{12} = p \vctr{e}_z$, thus reducing the
number of free parameters to three.  

\subsection{Hubbard model of the spin triangle: direct exchange}%

\label{sub:hcu3}

In this Subsection we give a brief  description of the Hubbard model
for a triangular molecule with $D\sub{3h}$ symmetry. In this model we
assume only direct coupling between the magnetic centers, thus no
bridge in-between. Even so, this simplified model catches the main
features of the effective spin Hamiltonian and gives the microscopic
mechanisms for  the spin-electric coupling. The Hamiltonian describing
the electrons in the triangular molecule reads
\begin{eqnarray}
  H_{H}&=&\left[\sum_{i,\sigma}c^{\dagger}_{i\sigma}(t+i\sigma
    \lambda\sub{SO})c_{i+1,\sigma}+h.c.\right]\nonumber\\ &+&\sum_{i,\sigma}\left(\epsilon_{0}n_{i\sigma}+\frac{1}{2}Un_{i\sigma}n_{i\bar{\sigma}}\right),
\label{hamSQ}
\end{eqnarray}    
where $\lambda\sub{SO}\equiv p = \vctr{P}_{ij} \cdot \vctr{e}_z$ is
the spin-orbit parameter (only one),  $\epsilon_0$ is the on-site
orbital energy, and $U$ is the on-site Coulomb repulsion energy. As
stated before, typically $\lambda\sub{SO}, |t|\ll U$,  which allows
for a perturbative treatment of the hopping and spin-orbit
Hamiltonians.  These assumptions agree well with the numerical
calculations performed in \cite{PKP06}.  

The perturbation theory program involves the unperturbed states of the
system. The first set of unperturbed states are the  one-electron
states
\begin{equation}
|\phi^{\sigma}_i\rangle=c^{\dagger}_{i\sigma}|0\rangle,
\label{one_elec_states}
\end{equation}
while the three-electron states split in two categories: (i) the site
singly occupied states 
\begin{equation}
|\psi_{k}^{\sigma}\rangle=\prod_{j=1}^3c^{\dagger}_{j\sigma_j}|0\rangle,
\label{singly_occupied}
\end{equation}
with $\sigma_j=\sigma$ for $j\neq k$ and $\sigma_j=\bar{\sigma}$, for
$j=k$, and (ii) the double-occupied sites
\begin{equation}
|\psi_{kp}^{\sigma}\rangle=c^{\dagger}_{k\uparrow}c^{\dagger}_{k\downarrow}c^{\dagger}_{p\sigma}|0\rangle,
\label{double_occupied}
\end{equation}
with $k=1,2,3$ and $p\neq k$.

The states in Eqs. (\ref{one_elec_states}), (\ref{singly_occupied}) and
(\ref{double_occupied}) are degenerate with energies $E=\epsilon_0$,
$E=3\epsilon_0$ and  $E=3\epsilon_0+U$, respectively.  Note that these
state are eigenstates of the Hamiltonian in Eq. (\ref{hamSQ}) only in
the absence of tunneling and SOI.

The above defined states are not yet adapted to the symmetry of the
system, i.e.  they are not basis states of the corresponding
irreducible representations of D$_{3h}$ point group.  Finding these
states is required by the fact that the symmetry of the molecule is
made visible through the hopping and SOI terms in the Hubbard
Hamiltonian. This is accomplished by  using projector
operators\cite{T06}.  We obtain for the one-electron symmetry adapted
states.
\begin{eqnarray}
|\phi_{A_1'}^{\sigma}\rangle&=&\frac{1}{\sqrt{3}}\sum_{i=1}^3|\psi_i^{\sigma}\rangle,\\
|\phi_{E_{\pm}'}^{\sigma}\rangle&=&\frac{1}{\sqrt{3}}\sum_{i=1}^3\epsilon_{1,2}^{i-1}|\psi_i^{\sigma}\rangle,\\
\end{eqnarray}
where $A_2'$ and $E_{\pm}'$ are one-dimensional and two-dimensional
irreducible representations in D$_{3h}$, respectively. Similarly, the
symmetry adapted states with the singly-occupied magnetic centers read:
\begin{eqnarray}
|\psi_{A_2'}^{1\sigma}\rangle&=&\frac{1}{\sqrt{3}}\sum_{i=1}^3|\psi_{i}^{\sigma}\rangle,\\ 
|\psi_{E_{\pm}'}^{1\sigma}\rangle&=&\frac{1}{\sqrt{3}}\sum_{i=1}^3\epsilon^{i-1}_{1,2}|\psi_i^{\sigma}\rangle,
\label{singly_occupied}
\end{eqnarray} 
while the symmetry adapted states of the doubly-occupied magnetic centers read:
\begin{eqnarray}
|\psi_{A_{1,2}^{'}}^{2\sigma}\rangle&=&\frac{1}{\sqrt{6}}\sum_{i=1}^3(|\psi_{i1}^{\sigma}\rangle\pm|\psi_{i2}^{\sigma}\rangle),\\ 
|\psi_{E_{\pm}^{'1}}^{2\sigma}\rangle&=&\frac{1}{\sqrt{6}}\sum_{i=1}^3\epsilon^{i-1}_{1,2}(|\psi_{i1}^{\sigma}\rangle+|\psi_{i2}^{\sigma}\rangle),\\ 
|\psi_{E_{\pm}^{'2}}^{2\sigma}\rangle&=&\frac{1}{\sqrt{6}}\sum_{i=1}^3\epsilon^{i-1}_{1,2}(|\psi_{i1}^{\sigma}\rangle-|\psi_{i2}^{\sigma}\rangle).
\label{doubly_occupied}
\end{eqnarray}

The tunneling and SOI mixes the singly-occupied and doubly-occupied
states.  Since both the tunneling and SOI terms in the Hubbard
Hamiltonian transform as the totally symmetric irreducible
representation $A_1'$ in D$_{3h}$, only states transforming according
to the same irreducible representations $\Gamma$ mix.  We obtain  the
perturbed in first order in $t/U$ and $\lambda\sub{SO}$:
\begin{eqnarray}
|\Phi_{A_2^{'}}^{1\sigma}\rangle&\equiv&|\psi_{A_2^{'}}^{1\sigma}\rangle,\\ 
|\Phi_{E_{\pm}^{'}}^{1\sigma}\rangle&\equiv&|\psi_{E_{\pm}^{'}}^{1\sigma}\rangle+\frac{(\bar{\epsilon}-1)(t\pm\sigma\lambda\sub{SO})}{\sqrt{2}U}|\psi_{E^{'1}_{\pm}}^{2\sigma}\rangle\nonumber\\
&+&\frac{3\epsilon(t\pm\sigma\lambda\sub{SO})}{\sqrt{2}U}|\psi_{E^{'2}_{\pm}}^{2\sigma}\rangle.
\label{renormalized_states}
\end{eqnarray}

\begin{figure}[t]
\begin{center}
\includegraphics[width=7cm]{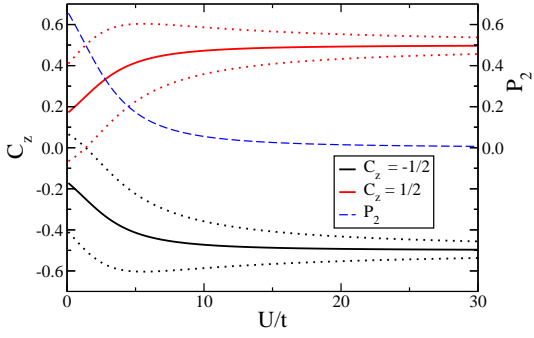}
\caption{
\label{fig:hubbardtospin}
Spin-Hamiltonian limit.  Expectation values of chirality $\langle C_z
\rangle$ (full lines) and the their bounds of uncertainty $\langle C_z
\rangle \pm \Delta C_z$ (dotted lines), see text, in the low-energy
states of the Hubbard model, as a function of the on-site repulsion
$U$, at the fixed hopping matrix element $t=1$ (left scale).  The
dashed line shows dependence of the double occupancy probability in
the ground state on the right scale.  The spin-Hamiltonian description
becomes accurate in the $U \rightarrow \infty$ limit.  The approach to
this limit is slow, and the double occupancy probability is
proportional to $t/U$.}
\end{center}
\end{figure}

Doubly occupied states become high in energy when
$|t|/U,\lambda\sub{SO}/U \ll 1$.  In this limit,  the orbital states
are quenched into singly-occupied localized atomic orbitals, and
low-energy behavior is determined by spin and described by a spin
Hamiltonian. In this limit the states in Eq. (\ref{singly_occupied})
are exactly the same chiral states in the spin Hamiltonian, i.e
$|\psi_{E_{\pm}'}^{1\sigma}\rangle\equiv|\psi_{\sigma}^{(1,2)}\rangle$
and $|\psi_{A_2'}^{1\sigma}\rangle\equiv|\psi_{\sigma}^{(0)}\rangle$.
The probability of finding two electrons at the same site decays as
$1/U$.  The lowest energy states have total spin $S=1/2$ and the
chirality $C_z = \pm 1$, and the fluctuations of chirality $\Delta C_z
= \sqrt{\langle C_z^2 \rangle - \langle C_z \rangle ^2}$ in the
eigenstates vanish, see Fig. {\ref{fig:hubbardtospin}}. 
The chiral states emerge as the eigenstates in the large-$U$ limit,
when the system is well described by the spin Hamiltonian.

The coupling of the molecule to an external electric field $\vctr{E}$
takes place via two mechanisms. The first one implies  modification of
the on-site single particle energies $\epsilon_0$ and leads to  the
following electric-dipole coupling Hamiltonian 
\begin{eqnarray}
H\sub{e-d}^0&=&-e\sum_{\sigma}\frac{E_ya}{\sqrt{3}}c^{\dagger}_{1\sigma}c_{1\sigma}-\frac{a}{2}\left(\frac{E_y}{\sqrt{3}}+E_x\right)c^{\dagger}_{2\sigma}c_{2\sigma}\nonumber\\ &+&\frac{a}{2}\left(E_x-\frac{E_y}{\sqrt{3}}\right)c^{\dagger}_{3\sigma}c_{3\sigma},
\end{eqnarray}
with $a$ being the geometrical distance between the magnetic ions and
$E_{x,y}$ the in-plane components of the electric field. The second
mechanism is due to modification of the hopping parameters $t_{ii+1}$
in electric field and gives 
\begin{equation}
H\sub{e-d}^1=\sum_{i,\sigma}t^{\vctr{E}}_{ii+1}c^{\dagger}_{i\sigma}c_{i+1\sigma},
\end{equation}   
where
$t^{\vctr{E}}_{ii+1}=\langle\Phi_{i\sigma}|-e\vctr{r}\cdot\vctr{E}|\Phi_{i+1\sigma}\rangle$
are new hopping parameters induced solely by the electric field
$\vctr{E}$, and $\Phi_{i\sigma}$ are the Wannier states localized on
the magnetic centers. We can write the $\vctr{E}$-induced hoppings as
$t^{\vctr{E}}_{ii+1}=\sum_{q=x,y,z}q_{ii+1}E_{q}$, with
$q_{ii+1}=\langle\Phi_{i\sigma}|-eq|\Phi_{i+1\sigma}\rangle$ being
electric dipole matrix elements  between the $i$ and $i+1$ ions. These
matrix elements are not all independent, symmetry alone reducing
drastically the  number of independent electric dipole parameters. In
order to find suitable independent free parameters, we switch from the
description in terms of localized  Wannier orbitals $\Phi_{i\sigma}$,
to the description in terms of  symmetry adapted states , namely from
$q_{ii+1}$ to
$q_{\Gamma\Gamma^{'}}=\langle\phi_{\Gamma\sigma}|q|\phi_{\Gamma^{'}\sigma}\rangle$,
where $\Gamma=A_1^{'},E_{\pm}^{'}$. In the basis of symmetry adapted
states, the components $q_{\Gamma\Gamma^{'}}$ satisfy a number of
relations.  In particular, we find:
\begin{eqnarray}
\langle\phi_{A_1'}^{\sigma}|-ex|\phi_{A_1'}^{\sigma}\rangle&=&\langle\phi_{A_1'}^{\sigma}|-ey|\phi_{A_1'}^{\sigma}\rangle=\langle\phi_{E_+'}^{\sigma}|-ex|\phi_{E_+'}^{\sigma}\rangle\equiv0\\ \langle\phi_{E_-'}^{\sigma}|-ex|\phi_{E_-'}^{\sigma}\rangle&=&\langle\phi_{E_+'}^{\sigma}|-ey|\phi_{E_+'}^{\sigma}\rangle=\langle\phi_{E_-'}^{\sigma}|-ey|\phi_{E_-'}^{\sigma}\rangle\equiv0,\\ \langle\phi_{E_+'}^{\sigma}|-ex|\phi_{E_-'}^{\sigma}\rangle&=&-i\langle\phi_{E_+'}^{\sigma}|-ey|\phi_{E_-'}^{\sigma}\rangle\equiv
d_{EE}\\ \langle\phi_{A_1'}^{\sigma}|-ex|\phi_{E_+'}^{\sigma}\rangle&=&\langle\phi_{A_1'}^{\sigma}|-ex|\phi_{E_-'}^{\sigma}\rangle
=-i\langle\phi_{A_1'}^{\sigma}|-ey|\phi_{E_+'}^{\sigma}\rangle\nonumber\\ &=&i\langle\phi_{A_1'}^{\sigma}|-ey|\phi_{E_-'}^{\sigma}\rangle\equiv
d_{AE}.
\end{eqnarray}
These relations reduce the number of free coupling constants to two,
namely $d_{EE}$ and $d_{AE}$.

It is instructive to write first the relation between the second
quantized operators $c^{\dagger}_{i\sigma}(c_{i\sigma})$  and
$c^{\dagger}_{\Gamma\sigma}(c_{\Gamma\sigma})$, which create
(annihilate) electrons in  localized and symmetry adapted states,
respectively:
\begin{equation}
\left(\begin{array}{c}
  c^{\dagger}_{1\sigma}\\c^{\dagger}_{2\sigma}\\c^{\dagger}_{2\sigma}
\end{array}
\right)=\frac{1}{\sqrt{3}}\left(
\begin{array}{ccc}
1 & 1 & \epsilon\\ 1 & \bar{\epsilon} & \bar{\epsilon}\\ 1 & \epsilon
& 1
\end{array}\right)
\left(\begin{array}{c}c^{\dagger}_{A_1'\sigma}\\c^{\dagger}_{E_+'\sigma}\\c^{\dagger}_{E_-'\sigma}\end{array}\right).
\label{symmetry_operators}
\end{equation}
With these expressions at hand,  we can write the electric dipole
Hamiltonian together with the spin-orbit Hamiltonian in the following
form:
\begin{eqnarray}
H\sub{e-d}^0&=&\frac{-iea\sqrt{3}}{2}\sum_{\sigma}\big(\bar{E}c_{E_+'\sigma}^{\dagger}c_{A_1'\sigma}-\epsilon Ec_{E_-'\sigma}^{\dagger}c_{A_1'\sigma}\nonumber\\
&+&\epsilon\bar{E}c_{E_-'\sigma}^{\dagger}c_{E_+'\sigma}\big)+H.c.,\\ 
H\sub{e-d}^1&=&\sum_{\sigma}d_{AE}(\bar{E}c^{\dagger}_{A_1'\sigma}c_{E_+'\sigma}-Ec^{\dagger}_{A_1'\sigma}c_{E_-'\sigma})\nonumber\\ &+&\bar{E}d_{EE}c^{\dagger}_{E_+'\sigma}c_{E_-'\sigma}+H.c.,\\ 
H\sub{SO}&=&\sqrt{3}\lambda\sub{SO}\sum_{\sigma}\sigma(c^{\dagger}_{E_-'\uparrow}c_{E_-'\uparrow}-c^{\dagger}_{E_+'\uparrow}c_{E_+'\uparrow}),
\end{eqnarray}
where $E=E_x+iE_y(\bar{E}=E_x-iE_y)$. The symmetry adapted states can
also be expressed  in terms of the symmetry adapted operators
$c^{\dagger}_{\Gamma}$. The expressions for these states are shown in
Appendix A. Using these states, we can compute all the matrix elements
corresponding to the electric dipole and SOI Hamiltonian,
respectively. The explicit form of these matrix elements can be found
in Appendix B. 

We now compute the electric dipole matrix elements between the
perturbed chiral states of the $E_{\pm}'$ symmetry. The question is to
what order in $t/U$ and/or $eEa(d_{EE},d_{AE})/U$ we want to do it. We
use the relations $|ea|\gg d_{EE}, d_{AE}$, which hold in  the
case of localized  orbitals.  This leads us to the following matrix
element of the electric dipole in the ground state:
\begin{eqnarray}
|\langle\Phi_{E_-^{'}}^{1\sigma}|H\sub{e-d}^0|\Phi_{E_+^{'}}^{1\sigma}\rangle|&\propto&\left|\frac{t^3}{U^3}eEa\right|,\\ 
|\langle\Phi_{E_-^{'}}^{1\sigma}|H\sub{e-d}^1|\Phi_{E_+^{'}}^{1\sigma}\rangle|&\simeq&\left|\frac{4t}{U}Ed_{EE}\right|.
\end{eqnarray}

We now relate the SOI matrix elements to the DM vectors in the
effective spin-Hamiltonian.  In $D_{3h}$ symmetry, the DM term reads
\begin{equation}
H\sub{SO}=\frac{iD_z}{2}\sum_{i=1}^{3}(S_+^{i}S_-^{i+1}-S_-^{i}S_+^{i+1}),
\end{equation}
which gives rise to  the following  non-zero matrix elements,
\begin{equation}
\langle\Phi_{E_{\pm}'}^{1\sigma}|H\sub{SO}|\Phi_{E_{\pm}'}^{1\sigma}\rangle=\pm\frac{\sqrt{3}D_z}{2}{\rm sign}(\sigma),
\end{equation}  
and allows us to make the following identification
\begin{equation}
D_z\equiv\frac{5\lambda\sub{SO}t}{U}.
\end{equation}
We see that this SOI term acts exactly as the 'microscopic' SOI
derived before: it splits the chiral states, but it does not mix
them. 

The Hubbard model with spin-orbit coupling can reproduce the
energy-level structure of the spin Hamiltonian.  In the limit of
strong on-site repulsion $|t|/U\ll 1$, the atomic orbitals in the
triangle vertices are occupied by one electron each.  The lowest
energy manifold consists of four states with the total spin
$S\sub{tot} = 1/2$.  These states are split from the next four-level
$S\sub{tot} = 3/2$ manifold by a gap of the order of $t^2/U$.

\subsection{Superexchange in molecular bonds}
\label{sub:bridge}

In this Subsection, we use the Hubbard model to deduce the dependence
of the spin Hamiltonian of MNs on the external electric fields in the
case where the coupling between magnetic sites is mediated by a 
non-magnetic bridge. In particular, we study how the parameters of the 
effective spin Hamiltonian depend on the hopping matrix elements that 
are modified by the presence of an electric field.
\begin{figure}[t]
\begin{center}
\includegraphics[width=8.5cm]{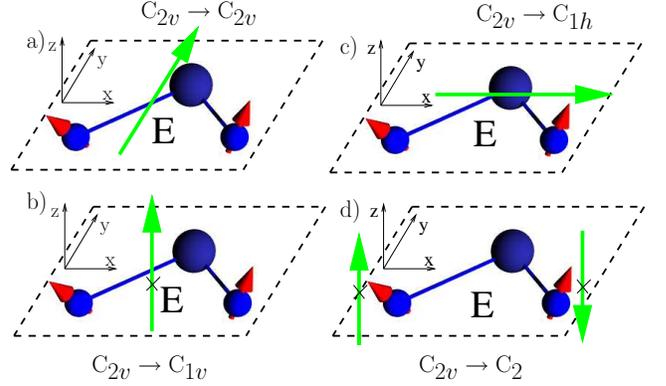}
\vspace{0.2cm}
\caption{\label{fig:bridge} Geometry of the bond and reduction of
  symmetry.  (a) Electric field $\vctr{E}$ in $y$ direction, leaves the ${\rm
    C}_{2v}$ symmetry unbroken. (b) An electric field $\vctr{E}$ in
$z$-direction, normal to the bond plane, reduces the symmetry to $\{ E
, \sigma_v \}$.  (c) An electric field $\vctr{E}$ in $x$-direction,
along the line connecting the magnetic centers,  reduces the symmetry to $\{ E,
  \sigma_h \}$.  (d) In an inhomogeneous staggered electric field
  $\vctr{E}$, the  reduced symmetry group is $\{ E, R_{y,\pi} \}$.
  } 
\end{center}
\end{figure}  
This method was successfully applied in the studies of strongly
correlated electrons, like cuprates \cite{YHE+94} and multiferroics
\cite{DYY+09}.

In order to describe the magnetic coupling, we consider a pair of sites
corresponding to the magnetic centers and a bridge site. Since the
direct overlap of the orbitals localized on the magnetic centers is
small, we set the direct hopping between the magnetic centers to zero,
but allow for the hopping of electrons between the magnetic sites and
the bridge site. This hopping gives rise to superexchange interaction
between the spins on the magnetic sites \cite{M60}. In
the limit of strong on-site repulsions, the effective Hamiltonian in
the lowest energy sector of the bond corresponds to a spin Hamiltonian 
where the coupling strengths are determined by the Hubbard model 
parameters. This correspondence provides an intuitive picture of the 
mechanism that leads to the interaction between the spins. It also allows us
to infer the properties of the molecule that lead to a strong
spin-electric coupling, e.g., the delocalization of the orbitals and
their local symmetry.

The Hubbard Hamiltonian of the bond is given by
\begin{equation}
\label{eq:hubbardhamiltonianbridge}
\begin{split}
H\sub{b} =&  \sum_{i,\alpha \beta} \left[ \cre{c}_{i\alpha} \left(
  t_i \delta_{\alpha \beta} + \frac{i\vctr{P}_{i}}{2} \cdot
  {\boldsymbol{\sigma}}_{\alpha \beta} \right) \ann{b}_{\beta} + \hc
  \right] \\ &+ U_1(n_{1}) + U_2(n_{2}) + U\sub{b}(n\sub{b}),
\end{split}
\end{equation}
where the indices $1$ and $2$ refer to the magnetic sites, and ${\rm
  b}$ refers to the bridge site.  We derive the spin Hamiltonian by
fourth-order Schrieffer-Wolff transformation of the Hamiltonian $H\sub{b}$ 
(\ref{eq:hubbardhamiltonianbridge}).

The Schrieffer-Wolf transformation \cite{W03} of the bond Hamiltonian
$H\sub{b} = H_0 + H\sub{tun}$ (\ref{eq:hubbardhamiltonianbridge}),
where the unperturbed Hamiltonian $H_0 = U_1(n_{1}) + U_2(n_{2}) +
U\sub{b}(n\sub{b})$ produces an effective low energy Hamiltonian
$H_{12}$ that approximately describes the low-energy dynamics of the
bond.  The effective Hamiltonian is
\begin{equation}
\label{eq:swtransformation}
H_{12} = \mathcal{P} \e{S} H\sub{b} \e{-S} \mathcal{P},
\end{equation}
where the antiunitary operator $S$ is chosen so that the low-energy
space of $H_0$ is decoupled from the high-energy space.  This operator
is found iteratively, $S= S^{(1)} + S^{(2)} + \ldots$, so that the nth
order transformation $S^{(n)}$ removes the terms that couple the low-
and high-energy states up to order $n$.  The projector $\mathcal{P}$
projects to the low-energy states.  In our system, the lowest order
Schrieffer-Wolff transformation that gives a nontrivial contribution
to the low-energy spin Hamiltonian is of  fourth order, and the
operator $S$ is approximated as $S \approx \sum_{n=1}^4 S^{(n)}$.

The unperturbed Hamiltonian,
$H_0= U_1 + U_2 + U\sub{b}$, describes localized electrons, and
the hopping $H\sub{tun}$ acts as perturbation.  The low-energy subspace of
the unperturbed Hamiltonian is spanned by the states in which the
magnetic ions are singly occupied, and the bridge is doubly occupied.
The lowest-order terms that give rise to a
nontrivial spin Hamiltonian, in the limit $ |t|, |\vctr{P}| \ll U $,
are of the fourth order in $t$ and $\vctr{P}$.  

The resulting interaction of the spins includes an isotropic exchange
of strength $J$, a Dzyalozhinsky-Moriya interaction described by a
vector $\vctr{D}$, and an anisotropic exchange term described by a
second rank symmetric traceless tensor ${\mathbf \Gamma}$
\cite{M60b} 
\begin{equation}
\label{eq:spinhamiltonianbridge}
H_{12} = J  \vctr{S}_1 \cdot \vctr{S}_{2}  + \vctr{D}
\cdot \left( \vctr{S}_1 \times \vctr{S}_{2} \right) + \vctr{S}_1 \cdot
      {\mathbf \Gamma}  \vctr{S}_{2}.
\end{equation}
Quite generally the interaction between two spins up to second order
in $\vctr{P}_{12}$ can be represented as an isotropic exchange of
rotated spins \cite{YHE+94}.  However, since the frustration in the
triangle is strong, it is a good approximation to take only the
Dzyalozhinsky-Moriya interaction into account for the weak spin-orbit
coupling, $|\vctr{P}_{12}| \ll |t_{12}|$ when describing a full
molecule. 

In a bond with a single bridge site, the largest possible symmetry is
${\rm C}_{2v}$.  We introduce Cartesian coordinates with the $x$-axis
pointing from the magnetic center $1$ to $2$, $y$-axis lying in the
bond plane and pointing towards the bridge site, and the
$z$-axis normal to the bond plane (Fig. \ref{fig:bridge}).  The
elements of the ${\rm C}_{2v}$ symmetry group
are then rotation $R_{y,\pi}$ by $\pi$ about the $y$-axis, reflection
$\sigma_v$ in the $yz$ plane, and reflection $\sigma_h$ in the $xy$
plane.  Each of these symmetry operations present imposes constraints
on the parameters of $H\sub{b}$.  In the case of
localized orbitals that remain invariant under the local symmetries of
their respective sites, the constraints resulting from the $R_{y,\pi}$
symmetry are:  
\begin{align}
\label{eq:rpiyconstraintst}
t_1 &= t_2, \\
\label{eq:rpiyconstraintsx}
P_{x,1} &= -P_{x,2}, \\
\label{eq:rpiyconstraintsy}
P_{y,1} &= P_{y,2}, \\
\label{eq:rpiyconstraintsz}
P_{z,1} &= -P_{z,2}.
\end{align}
The $\sigma_v$ symmetry implies:
\begin{align}
\label{eq:sigmavconstraintst}
t_1 &= t_2, \\
\label{eq:sigmavconstraintsx}
P_{x,1} &= P_{x,2}, \\
\label{eq:sigmavconstraintsy}
P_{y,1} &= -P_{y,2}, \\
\label{eq:sigmavconstraintsz}
P_{z,1} &= -P_{z,2},
\end{align}
and the $\sigma_h$ symmetry implies:
\begin{equation}
\label{eq:sigmahconstraints}
\vctr{P}_1=-\vctr{P}_2 = p\vctr{e}_z.
\end{equation}
In the perturbative calculation of the effective spin Hamiltonian
parameters, these constraints reproduce the Dzyalozhinsky-Moriya
rules. We do not deal with the symmetry of
on-site energies $U_{1,2,{\rm b}}$ in any detail, since they do not
affect the spin Hamiltonian at this level of approximation.

\subsection{Electric field along $y$}

In the electric field pointing along the $y$ axis, the point group
symmetry of the bridge remains ${\rm{C}}_{2v}$, and all of the
constraints (\ref{eq:rpiyconstraintsx}) --
(\ref{eq:sigmahconstraints}) hold.  The fourth-order Schrieffer-Wolff
transformation then gives the interaction between the spins on
magnetic centers of the form (\ref{eq:spinhamiltonianbridge}) with the
parameters 
\begin{align}
\label{eq:c2vj}
J &= \frac{1}{12 U^3}\left( 48 t^4 - 40 t^2 p_z^2 + 3 p_z^4 \right),
\\
\label{eq:c2vd}
\vctr{D} &= \frac{2}{U^3} t p_z \left( 4 t^2 - p_z^2 \right)
\vctr{e}_z, \\
\label{eq:c2vgamma}
\Gamma_{xx} &= \Gamma_{yy} = - \frac{1}{2} \Gamma_{zz} =
-\frac{8}{3U^3} t^2 p_z^2,
\end{align}
while all the off-diagonal elements of $\vctr{\Gamma}$ vanish.  Here,
the parameters of the Hubbard model satisfy the symmetry constraints
of the full ${\rm{C}}_{2v}$, and 
\begin{align}
\label{eq:c2vtparam}
t_1 &= t_2 = t, \\
\label{eq:c2vpzparam}
\vctr{P}_1 &= - \vctr{P}_2 = p_z \vctr{e}_z.
\end{align}
We have introduced $U^3 = U\sub{c2} ( 2 U\sub{c2} -
U\sub{b2} ) ( U\sub{b1} - U\sub{b2} + U\sub{c2} )^2 / ( 4 U\sub{c2} -
U\sub{b2} )$, where the on-site repulsions are $U\sub{b2}$ for the doubly
occupied bridge, $U\sub{b1}$ for the singly occupied bridge, and
$U\sub{c2}$ for the doubly occupied magnetic center.  The parameter $U$
describes the energy cost of leaving the manifold of states with the
minimal energy of Coulomb repulsion.  We assume that
the lowest energy charge configuration corresponds to a doubly
occupied bridge, so that $U\sub{b2}<U\sub{b1}$. 

In  first order, the variations of the spin-Hamiltonian parameters
resulting from the modification of the Hubbard model parameters, are:
\begin{align}
\label{eq:c2vdeltaj}
\delta J &= \frac{1}{3 U^3} \left[ \left( 48 t^3 - 20 t p_z^2 \right)
  \delta t + \left( - 20 t^2 p_z + 3 p_z ^3 \right) \delta p_z
  \right],  \\
\label{eq:c2vdeltadz}
\delta D_z &= \frac{2}{U^3} \left[ \left( 12 t^2 p_z - p_z^3 \right)
  \delta t + \left( 4 t^3 - 3 t p_z^2 \right) \delta p_z \right], \\
\label{eq:c2vdeltagamma}
\delta \Gamma_{xx} &= \delta \Gamma_{yy} = -\frac{\delta \Gamma_{zz}}{2} 
 = -\frac{16t p_z }{3 U^3} \left( p_z \delta t + t \delta
p_z \right).
\end{align}
Electric field modifies the orbitals and therefore the overlaps 
between them, that determine the hopping parameters.  We consider the
case where the variations $\delta t$ and $\delta p_z$ are linear in
the field intensity $ E_y $:  $ \delta t   = \kappa_{t  } E_y $, 
$ \delta p_z = \kappa_{p_z} E_y$.  We will not discuss the effect of
variations in the on-site energies  $U$ in any details, since their
only effect in the fourth order perturbation is a rescaling of all the
spin Hamiltonian parameters by  $ U^3 / ( U+ \delta U)^3$.

We stress that these linear modifications of the hopping parameters
are  characteristic for the ${\rm C}_{2v}$ symmetry.  If the electric
field is oriented differently and thus lowers the system symmetry (see
below) first-order  increments are not allowed, and the spin-electric
coupling is at least a second order effect in the electric field.  The
modification of the orbitals includes the energy scale of splitting of
the atomic orbitals in the molecular field. We have assumed earlier
that the splitting of the orbitals localized on the magnetic centers
is large, and the dominant source of the spin-electric coupling is the
modification of the bridge orbital.  Therefore, the key criterion for
strong spin-electric coupling is the presence of bridge orbitals that
are weakly split in the molecular field.   If, in addition, we assume
that the modification is a property of the bond alone, and not of the
entire molecule, the $\kappa$ parameters can be determined in an
ab-initio calculations on a smaller collection of atoms.  

In the limit of weak spin-orbit coupling, $|t| \gg |p_z|$, the main
effect of the electric fields is a change of $J$, leading to our
symmetry-based results, see Eq.\ref{full_effective}.  In particular,
the $d$ parameter of the symmetry analysis is:
\begin{equation}
\label{eq:dofalpha}
\begin{split}
d = \frac{4}{U^3} & \left[ \left( 48 t^3 - 20 t p_z^2 \right)
  \kappa_{t} + \right.  \\ & \quad \left. \left( - 20 t^2 p_z + 3 p_z
  ^3 \right) \kappa_{pz} \right] . 
\end{split}
\end{equation}

In this case, the Dzyalozhinsky-Moriya vector $\vctr{D}$ is constraint
to point in the $z$ direction, $\vctr{D}=D\vctr{e}_z$.  The model
suggests that the dominant effect of the electric field in the
molecules with dominant Heisenberg exchange ($J \gg |\vctr{D}|$) is
modification of the isotropic exchange constants $J$, and 
\begin{equation}
\label{eq:bridgec2v}
\frac{|\delta \vctr{D}|}{|\delta J|} \sim \frac{|\vctr{D}|}{|J|}, 
\end{equation}
so that the modification of the Dzyalozhinsky-Moriya vector $\vctr{D}
\rightarrow \vctr{D} + \delta \vctr{D}$ is weaker.  However, in the
molecules in which the modifications of $J$ are inefficient in
inducing the spin-electric coupling, as for example in the spin-$1/2$
pentagon, the modifications of $\vctr{D}$ may eventually provide the
main contribution to the spin-electric coupling. 

Electric field pointing in a generic direction breaks the ${\rm
  C}_{2v}$ symmetry of the bridge, and allows further modification of
the Hubbard and spin Hamiltonian parameters, that do not obey all the
symmetry constraints in Eqs.  (\ref{eq:rpiyconstraintst}) --
(\ref{eq:sigmahconstraints}).  With the relaxed constraints, both the
direction and intensity of $\vctr{P}_{1,2}$, as well as the
spin-independent hoppings $t_{1,2}$ become field-dependent.  This
observation can be used in the search for molecules that show strong
spin-electric coupling.  The energy cost of changing the distance
between the localized orbitals may be significantly higher than the
cost of modifying the shape of the bridge orbital.  In order to
investigate this dependence, we study the effective spin Hamiltonian
description of a bridge with all possible residual symmetries.  

\subsubsection{Residual $\sigma_v$ symmetry}
\label{sec:subsubsigmav}

An electric field $\vctr{E} = E \vctr{e}_z$ normal to the bond's plane
reduces the initial ${\rm C}_{2v}$ symmetry down to $\left\{ E,
\sigma_v\right\}$.  This reduction of the symmetry also happens when a
molecule is deposited on the surface parallel to the bond plane.
While the constraints in Eq. \ref{eq:sigmahconstraints} hold, this
reduction of symmetry implies the appearance of nonzero in-plane
components of $\vctr{P}_{1,2}$.  We parameterize the most general
Hubbard model parameters $t_{1,2}$, $\vctr{P}_{1,2}$ consistent with
the symmetry as
\begin{align}
\label{eq:sigmaht}
t_1 &= t_2 = t, \\
\label{eq:sigmahpx}
P_{1,x} &= P_{2,x} = p_{xy} \cos{\phi}, \\
\label{eq:sigmahpy}
P_{1,y} &= -P_{2,y} = p_{xy} \sin{\phi}, \\
\label{eq:sigmahpz}
P_{1,z} &= -P_{2,z} = p_z.
\end{align}
The effective low energy spin Hamiltonian, derived by Schrieffer-Wolff
transformation up to fourth order in $t/U$, and $|\vctr{P}|/U$ is
given by (\ref{eq:spinhamiltonianbridge}), with the non-zero
parameters
\begin{widetext}
\begin{align}
\label{eq:jsigmav}
J &= \frac{1}{12 U^3}  \left[ p_{xy}^4 - 2 p_{xy}^2 p_{z}^2 + 3
p_{z}^4 - 8 t^2 \left( p_{xy}^2 + 5 p_{z}^2 \right) + 48 t^4 - 8
p_{xy} ^2 \left( p_{z}^2 - 4 t^2 \right) \cos{2 \phi} + 2 p_{xy}^4
\cos{4 \phi} \right], \\
\label{eq:dysigmav}
D_{y} &= -\frac{p_{xy}}{U^3}  \left( p_{z} \cos{\phi} + 2 t
\sin{\phi} \right) \left( - p_{z}^2 + 4 t^2 + p_{xy}^2 \cos{2\phi}
\right), \\
\label{eq:dzsigmav}
D_{z} &= -\frac{1}{2 U^3} \left( 4 t p_{z} - p_{xy}^2 \sin{2\phi}
\right) \left( p_{z}^2 - 4 t^2 - p_{xy}^2 \cos{2\phi} \right), \\
\label{eq:gammaxxsigmav}
\Gamma_{xx} &= -\frac{1}{6 U^3} \left[ p_{xy}^2 \left( 1 - \cos{2\phi}
\right) + 2 p_{z}^2 \right] \left[ 8 t^2 + p_{xy}^2 \left( 1 +
\cos{2\phi} \right) \right], \\
\label{eq:gammayysigmav}
\Gamma_{yy} &= \frac{1}{12 U^3}  \left\{ - p_{xy}^4 + 8 p_{xy}^2
p_{z}^2 + 32 t^2 \left( p_{xy}^2 - p_{z}^2 \right) + p_{xy}^2 \left[ 8
\left( p_{z}^2 - 4 t^2 \right) \cos{2\phi} + p_{xy}^2 \cos{4\phi} + 48
t p_{z} \sin{2\phi} \right] \right\}, \\
\label{eq:gammayzsigmav}
\Gamma_{yz} &= \Gamma_{zy} = \frac{p_{xy}}{U^3} \left( p_{z}\cos{\phi} + 2 t
\sin{\phi} \right) \left( - 4 t p_{z} + p_{xy}^2 \sin{2\phi} \right)
\\
\label{eq:gammazzsigmav}
\Gamma_{zz} &= - \Gamma_{xx} - \Gamma_{yy}.
\end{align}
\end{widetext}
In the lowest order in spin-orbit coupling the spin interaction
consists of the isotropic exchange with $J\approx 4 t^4/U^3$, and the
Dzyalozhinsky-Moriya interaction with $\vctr{D}\approx -8t^3(
p_{xy}\sin{\phi} \vctr{e}_y + p_{z} \vctr{e}_z)/U^3$.  

As a matter of principle, the spin-orbit coupling mediated hopping
$\vctr{P}$ does not have to be much weaker than the spin-independent
hopping $t$.  In this case, all the nonzero terms in Eqs.
(\ref{eq:jsigmav}) --- (\ref{eq:gammazzsigmav}) are of comparable
size, and the variation of spin Hamiltonian with the angle $\phi$
becomes significant.  Note that the angle $\phi$ describes the
directions of spin-orbit coupling induced hopping parameters
$\vctr{P}_{1,2}$, and that it is not directly connected to the bond
angle between the magnetic sites and the bridge site.  However, for
the bridge orbital without azimuthal symmetry, the angle $\phi$ does
depend on the bond angle.  For the molecules in which the full
symmetry allows only for the spin-electric coupling mediated by the
spin-orbit interaction, this effect is important. 

With these assumptions, the dependence of the effective spin
Hamiltonian on $p_{xy}$ suggests that the strength of induced in-plane
Dzyalozhinsky-Moriya vector will be sensitive to the angle $\phi$ that
is determined by the angular dependence of the bridge- and magnetic
center orbitals.  In turn, for a fixed symmetry of the bridge orbital,
this dependence directly translates into the dependence of the
spin-electric coupling constant on the bridge bond angle.

In the presence of electric field $\vctr{E} = E \vctr{e}_z$, the
hopping parameters will change from their initial values, that satisfy
the constraints implied by the ${\rm C}_{2v}$ symmetry, into a set of
values that satisfy those implied by $\sigma_v$ only.  The resulting
change in the spin-Hamiltonian parameters reads:
\begin{align}
\label{eq:sigmavdeltaj}
\delta J &= \frac{1}{3 U^3} \left[ 4 t_0 \left( 12 t_0^2 - 5 p_{z0}^2
  \right) \delta t \right.  \\ \nonumber & \qquad \left.  + p_{z0}
  \left( - 20 t_0^2 + 3 p_{z0}^2 \right) \delta p_{z} \right], \\
\label{eq:sigmavdeltady}
\delta D_y &= -\frac{1}{U^3} \left( 4 t_0^2 - p_{z0}^2 \right) \left(
2 t_0 \sin{\phi} + p_{z0} \cos{\phi} \right) \delta p_{xy}, \\
\label{eq:sigmavdeltadz}
\delta D_z &= \frac{2}{U^3} \left[ p_{z0} \left( 12 t_0^2 - p_{z0}^2
  \right) \delta t \right.  \\ \nonumber & \qquad \left.  + t_0 \left(
  4 t_0^2 - 3 p_{z0}^2 \right) \delta p_z \right], \\
\label{eq:sigmavdeltagammaxxyyzz}
\delta \Gamma_{xx} &= \delta \Gamma_{yy} = -\frac{1}{2} \delta
\Gamma_{zz} =  \\ \nonumber & \qquad  - \frac{16}{3U^3} t_0 p_{z0}
\left( p_{z0} \delta t + t_0 \delta p_z \right),  \\
\label{eq:sigmavdeltagammayz}
\delta \Gamma_{yz} &= \delta \Gamma_{zy} = - \frac{4}{U^3} t_0 p_{z0}
\left( 2 t_0 \sin{\phi} + p_{z0} \cos{\phi} \right) \delta p_{xy}.
\end{align}

The $\sigma_v$-symmetric variations of Hubbard parameters occur when an
external electric field is applied along the $z$ direction to a ${\rm
  C}_{2v}$ symmetric bond.  Again, the variations of the parameters is
generically linear in the field strength, $\delta t = \kappa_{t,\sigma v} E_z$, $\delta
p_{xy} = \kappa_{pxy,\sigma v} E_z $, $\delta p_z = \kappa_{pz, \sigma
  v} E_z$, where the $\kappa$ parameters depend on the modification of
the bridge orbital in the electric field.  As opposed to the case of
the field along $y$ direction that maintains the bonds ${\rm C}_{2v}$
symmetry, the $\kappa$ parameters for the field along $z$ axis vanish
in zero field, since the $z$-component of a vector has no matrix
elements between the relevant ${\rm C}_{2v}$-symmetric states.  The
linear expansion is valid when the field is strong enough to distort
the bridge orbital.  Alternatively, the expansion is valid for a bond
with lower symmetry in zero electric field, e. g., when the bond is close to a surface.

\subsubsection{Residual $\sigma_h$ symmetry}

In an electric field that lies in plane of the bond, with $\vctr{E}
\parallel \hat{\vctr{x}}$), the only residual symmetry transformation
is the reflection about the $xy$ plane ($ \sigma_h $).  Within this
reduced symmetry, the two magnetic sites are no longer equivalent, but
the spin-dependent hopping parameters $\vctr{P}_{1,2}$ still point
along the $z$ axis:
\begin{align}
t_1 \neq t_2,  \, \vctr{P}_1 = p_1 \vctr{e}_z \neq p_2 \vctr{e}_z =
\vctr{P}_2 .
\end{align}

In the fourth order in hopping $t$, $\vctr{P}$, the resulting low
energy spin Hamiltonian is again given by
Eq. \ref{eq:spinhamiltonianbridge}, with the following non-zero
coupling constants:
\begin{align}
\label{eq:jsigmah}
J &= \frac{1}{12 U^3} \left[ 32 t_{1} t_{2} p_{1z} p_{2z} \right.
\\ \nonumber & \left. -  4 \left( t_{1}^2 p_{2z}^2 + t_{2}^2 p_{1z}^2
\right) + 48 t_{1}^2 t_{2}^2 + 3 p_{1z}^2p_{2z}^2 \right], \\
\label{eq:dsigmah}
\vctr{D} &= -\frac{1}{U^3} \left( t_{1} p_{2z} - t_{2} p_{1z} \right)
\left( 4 t_{1} t_{2} + p_{1z} p_{2z} \right) \vctr{e}_z, \\
\label{eq:gammasigmah}
\Gamma_{xx} &= \Gamma_{yy} = -\frac{\Gamma_{zz}}{2} = -\frac{2}{3
  U^3} \left( t_{1} p_{2z} - t_{2} p_{1z} \right)^2,
\end{align}
Similarly to the case of full ${\rm C}_{2v}$ symmetry, the spin
Hamiltonian consists of the isotopic exchange $J$, Dzyalozhinsky-Moriya
vector $\vctr{D}=D_z {\hat{\vctr{z}}}$ normal to the bond plane, and
diagonal tensor $\Gamma$ isotropic in the bond plane ($\Gamma_{xx} =
\Gamma_{yy}$).  We stress that the dependence of the effective spin
Hamiltonian parameters on those entering the spin Hubbard Hamiltonian
is  different for these two symmetries, and so is the response to the
applied electric field.  On one hand, the ${\rm C}_{2v}$ preserving
electric field induces the transitions in the lowest energy multiplet
in the lowest order.  On the other hand, the electric field that
reduces the bond symmetry to $\{ E, \sigma_h \}$ does not alter the
coupling of spins in the lowest order, since the deformation of the
molecule requires some coupling to the field.  

As in previous case, we expand the $\sigma_h$ symmetric spin
Hamiltonian around the ${\rm C}_{2v}$ symmetric case.  We introduce a
perturbation of the parameters Hubbard parameters in the electric
field consistent with the residual symmetry: 
$ t_1 = t_0 + \delta t_1 $, $ t_2 = t_0 + \delta t_2 $, 
$ p_{1z} =   p_{z0} + \delta p_{1z} $, 
$ p_{2z} = - p_{z0} + \delta p_{2z} $. 
As a consequence, the spin Hamiltonian parameters are incremented by:
\begin{align}
\label{eq:sigmahdeltaj}
\delta J =& \frac{1}{6 U^3} \left[ 4 t_0 \left( 12 t_0^2 - 5 p_{z0}^2
  \right) \left( \delta t_1 + \delta t_2 \right) \right.  \\ \nonumber
  & \left.  + p_{z0} \left( - 20 t_0^2 + 3 p_{z0}^2 \right) \left(
  \delta p_{1z} - \delta p_{2z} \right) \right], \\
\label{eq:sigmahdeltadz}
\delta D_z =& \frac{1}{U^3} \left[ p_{z0} t_0 \left( 12 t_0 - p_{z0}
  \right) \left( \delta t_1 + \delta t_2 \right) \right.  \\ \nonumber
  & \left.  + t_0 \left( 4 t_0^3 - 3 p_{z0}^2 \right) \left( \delta
  p_{1z} - \delta p_{2z} \right) \right], \\
\label{eq:sigmahdeltagxxgyygzz}
\delta \Gamma_{xx} =& \delta \Gamma_{yy} = -\frac{\delta\Gamma_{zz} }{2} 
=  \\ \nonumber & -\frac{8}{3U^3}  t_0 p_{z0} \left[
  p_{z0} \left( \delta t_1 + \delta t_2 \right) + t_0 \left( \delta
  p_{1z} - \delta p_{2z} \right) \right].
\end{align}
As for the case of $\sigma_v$ residual symmetry, there is no
spin-electric effect of the first order in electric field, and the
crucial condition for coupling to the electric field in this direction
is weak splitting of the bridge orbitals in the molecular field.

\subsubsection{Residual $R_{y,\pi}$ symmetry}
 
Reduction of the symmetry of the bond, from the full ${\rm C}_{2v}$ to
the group $\{ E, R_{y,\pi} \}$, does not occur for any vector
perturbation.  In terms of electric fields, this reduction of the
symmetry would correspond to an inhomogeneous electric field that
points in the $\vctr{e}_z$ direction at the position of one of the
magnetic centers, and in the $-\vctr{e}_z$ direction at the position
of the other.  This symmetry breaking can also happen when the localized
orbitals on the magnetic centers have lobes of opposite signs
extending in the $z$-direction, and oriented opposite to each other.  

The most general Hubbard model parameters consistent with the residual
symmetry are 
\begin{align}
t_1 &= t_2 = t, \\ P_{1x} &= - P_{2x} = p_{xy} \cos{\phi}, \\ P_{1y}
&= P_{2y} = p_{xy} \sin{\phi}, \\ P_{1z} &= P_{2z} = p_z.
\end{align}

After the fourth-order Schrieffer-Wolff transformation, the effective
low-energy spin Hamiltonian has the form
(\ref{eq:spinhamiltonianbridge}) with nonzero parameters
\begin{widetext}
\begin{align}
\label{eq:jrpiy}
J &= \frac{1}{12 U^3} \left( p_{xy}^4 - 2 p_{xy}^2 p_{z}^2 + 3 p_{z}^4
- 8 t^2 \left( p_{xy}^2 + 5 p_{z}^2 \right) + 48 t^2 + 8 p_{xy}^2
\left( p_{z}^2 - 4 t^2 \right) \cos{2\phi} + 2 p_{xy}^4 \cos{4\phi}
\right), \\
\label{eq:dxrpiy}
D_{x} &= \frac{1}{U^3} p_{xy} \left( - 2 t \cos{\phi} + p_{z}
\sin{\phi} \right) \left( p_{z}^2 - 4 t^2 + p_{xy}^2 \cos{2\phi}
\right), \\
\label{eq:dzrpiy}
D_{z} &= -\frac{1}{2 U^3} \left( 4 t p_{z} + p_{xy}^2 \sin{2\phi}
\right) \left( p_{z}^2 - 4 t^2 + p_{xy}^2 \cos{2\phi} \right), \\
\label{eq:gammaxxrpiy}
\Gamma_{xx} &= \frac{1}{12 U^3}  \left( - p_{xy}^4 + 8 p_{xy}^2
p_{z}^2 + 32 t^2 \left( p_{xy}^2 - p_{z}^2 \right) + p_{xy}^2 \left(
-8 \left( p_{z}^2 - 4 t^2 \right) \cos{2\phi} + p_{xy}^2 \cos{4\phi} -
48 t p_{z} \sin{2\phi} \right) \right), \\
\label{eq:gammaxzrpiy}
\Gamma_{zx} &= \Gamma_{xz} = \frac{1}{U^3} p_{xy} \left( 2 t \cos{\phi} - p_{z}
\sin{\phi} \right) \left( 4 t p_{z} + p_{xy}^2 \sin{2\phi} \right), \\
\label{eq:gammayyrpiy}
\Gamma_{yy} &= \frac{1}{6 U^3}  \left( p_{xy}^2 \left( 1 + \cos{2\phi}
\right) + 2 p_{z}^2 \right) \left( p_{xy}^2 \left( - 1 + \cos{2\phi}
\right) - 8 t^2 \right), \\
\label{eq:gammazzrpiy}
\Gamma_{zz} &= -\Gamma_{xx} - \Gamma_{yy} = - \frac{1}{6 U^3} \left(
-p_{xy}^4 + 2 p_{xy}^2 p_{z}^2 + 8 t^2 \left( p_{xy}^2 - 4 p_{z}^2
\right) + p_{xy}^2 \left( - 2 \left( p_{z}^2 - 4  t^2 \right)
\cos{2\phi} + p_{xy}^2 \cos{4\phi} - 24 t p_{z} \sin{2\phi} \right)
\right).
\end{align}
\end{widetext}

The expansion from the ${\rm C}_{2v}$ symmetric case gives (see the
discussion of the $\sigma_v$ residual symmetry in Subsection
\ref{sec:subsubsigmav}): 
\begin{align}
\label{eq:rpiydeltaj}
\delta J =& \frac{1}{3U^3} \left[ 4 t_0 \left( 12 t_0^2 - 5 p_{z0}^2
  \right) \delta t + p_{z0} \left( -20 t_0^2 + 3 p_{z0}^2 \right)
  \delta p_z \right], \\
\label{eq:rpiydeltdx}
\delta D_x =&  \frac{1}{U^3} \left( 4 t_0^2 - p_{z0}^2 \right) \left(
2 t_0 \cos{\phi_0} - p_{z0} \sin{\phi_0} \right) \delta p_{xy}, \\
\label{eq:rpiydeltadz}
\delta D_z =& \frac{2}{U^3} \left[ p_{z0} \left( 12 t_0^2 - p_{z0}^2
  \right) \delta t + t_0 \left( 4 t_0^2 - 3 p_{z0}^2 \right) \delta
  p_{z0} \right], \\
\label{eq:rpiydeltagxxgyygzz}
\delta \Gamma_{xx} =& \delta \Gamma_{yy} = -\frac{1}{2} \delta
\Gamma_{zz} =  -\frac{16}{3U^3} p_{z0} t_0 \left( p_{z0} \delta t +
t_0 \delta p_z \right), \\
\label{eq:rpiydeltagzx}
\delta \Gamma_{zx} =& \delta \Gamma_{xz} = \frac{4}{U^3} t_0 p_{z0} \left( 2 t_0
\cos{\phi_0} - p_{z0} \sin{\phi_0} \right) \delta p_{xy}.
\end{align}

As in the case of $\sigma_v$ symmetry, the resulting interaction of
the spins on magnetic centers becomes dependent on the angle $\phi$
between the two $\vctr{P}$ parameters.  This dependence is pronounced
in the case of strong spin-orbit coupling and can lead to the
dependence of spin-electric effects on both the geometry of the bond
and the shape of the bridge orbital.

\subsection{Bond modification and symmetries}

Spin-electric coupling induced by the superexchange through bridge
atoms depends on the symmetry of the bridge and the direction of the
electric field.  This symmetry reflects on the resulting coupling of
spins in an MN.  In this subsection, we combine the results of the
Hubbard model study of the individual bonds with the previous symmetry
considerations, and provide  rough estimates of the most promising
spin-electric coupling mechanism in the triangular and pentagonal
molecules.

The spin-electric coupling via superexchange is most
sensitive to the electric fields that does not break the initial ${\rm
  C}_{2v}$ local symmetry of the bond.  This symmetry corresponds to
the electric field that lies in the plane of the molecule and normal
to the bond.  All the other couplings require modification of the
bridge orbitals, and are suppressed by a factor $d|\vctr{E}|/U\sub{d}$,
where $U\sub{d}$ is on-site repulsion on the bridge.
Assuming that this repulsion is strong, we can model the
spin electric coupling as a set of modifications of the spin
interactions $\delta H_{jj+1}$ between the neighboring magnetic
centers, with $||\delta H_{jj+1}||\propto |E_{\perp}\subup{bond}|$,
where $ E_{\perp}\subup{bond}$ is the projection of the electric field
normal to the bond and lying in the molecule's plane.

In the triangle, the strongest effects of electric field is
modification of exchange couplings $\delta J_{jj+1} = \delta J_0
\cos{(2j\pi/3 + \theta_0)}$, where the angle $\theta_0$ describes the
orientation of the in-plane component of the electric field, and
$\delta J_0$ is a molecule-dependent constant.  This modification
leads to a specific coupling of the in-plane components of chirality
to the electric field $H\sub{e-d}\subup{eff} = d \vctr{E}' \cdot
\vctr{C}_{\parallel}$, see (\ref{effective_edipole}).  Other types of coupling are suppressed either
due to weaker influence of electric field on the bonds, or due to the
symmetry of the molecule.  If the spin-electric coupling is mediated
by the spin-orbit interaction, the suppression is by a factor of the
order $|\vctr{D}|/J$, and if the coupling is mediated by electric field, the
suppression factor is $d|\vctr{E}|/J$.  Assuming the simplest case,
the modification of exchange coupling is the most promising mechanism
for spin-electric coupling in triangular molecules.

In the pentagons, the modification of spin-spin interaction $\delta
H_{jj+1}$ preferred by the superexchange mechanism is inefficient in
inducing the spin-electric coupling of the molecule.  The pattern
$\delta J_{jj+1}$ of exchange coupling constants induced by an
external electric field does not couple the states within the lowest
energy manifold.  In order to couple the spins in the pentagon to an
external field, another mechanism is needed.  The modification of the
Dzyalozhinsky-Moriya vectors $\delta D_{jz} = \delta D_{z0} \cos{(2j\pi/5 +
  \theta_0)}$, where $\delta D_{z0}$ is a molecule-dependent constant, and
$\theta_0$ describes the orientation of the in-plane component of the
electric field, is preferred by the superexchange bridge model.  In
the symmetry analysis, we have found that this form of modification of
spin-orbit coupling does not induce  spin-electric coupling.  The
same applies to the modifications of in-plane components $\vctr{D}_{j,xy}$.  The
main effect that gives rise to spin-electric coupling is the
modification of the exchange interactions $\delta J_{jj+1}$ in the
presence of the original spin-orbit interaction $D_{jj+1,z}$.  Compared to a triangle
composed out of identical bonds, this interaction will be weaker by a
factor of ${|\vctr{D}_{jj+1}|}/J_{jj+1}$.

In summary, within our model of the superexchange-mediated
spin-electric coupling, the most promising candidates for the spin
manipulation via electric field are triangular molecules.  In
pentagons, the best candidates are molecules with strong spin-orbit
interaction, and weakly split bridge orbitals.

\section{Experimental signatures of the spin-electric coupling}
\label{sec:experiments}

Coherent quantum control of spins in an MN using electric fields can
be achieved by resonant driving of the transitions between the
chirality eigenstates \cite{TTS+08}.  At present, however, little is
known about the effects of electric fields on the  spin states of
molecular magnets.  As a preliminary step, it is useful to identify
possible signatures of such a coupling that are observable in  the
experiments routinely used  to characterize these systems.  

In this section, we study the ways in which the spin-electric coupling
can be detected in electron spin resonance (ESR), in nuclear magnetic
resonance (NMR), and in the thermodynamic measurement of an MN. 

\subsection{Electron spin resonance}\label{ssEPR}

Electron spin resonance (ESR) investigates transitions between states
belonging to a given $S$ multiplet and having different spin
projections $M$  along the magnetic field direction \cite{BG89}.
This technique provides information on the anisotropies of the spin
system, as well as on the chemical environment, and the spin dynamics
\cite{ARM+07}.  In the following, we show how the effects of an
external electric field can  show up in the ESR spectra of
antiferromagnetic spin rings by affecting both  the frequency and the
oscillator strength of the transitions.

\subsubsection{Triangle of $s=1/2$ spins}

We start by considering the simplest case of interest, namely that of
a  triangle of $s=1/2$ spins with $ {\rm D}_{3h}$ symmetry.  The
lowest energy eigenstates of the spin triangle, given in
Eq. (\ref{spin_states}) form an $S=1/2$.  The effective Hamiltonian
$H\sub{eff}$ of the molecule in the presence of electric and magnetic
field, and acting within this quadruplet is given by
Eq. (\ref{full_effective}).

We first consider the case of a static magnetic field perpendicular to
the molecule's plane (${\bf B}\parallel \hat{\bf z}$).  The
eigenvalues of $ H\sub{eff} $ are then given by:
\begin{eqnarray}\label{epreq1}
\lambda_{\sigma}^{\alpha} & = & \sigma [ \mathcal{B} + \alpha
  (\Delta\sub{SO}^2 + \mathcal{E}^2)^{1/2} ], 
\end{eqnarray}
where $\mathcal{E}\equiv d|\vctr{E}\times \hat{\vctr{z}}|$,
$\mathcal{B}=\mu_B\sqrt{g_{\parallel}^2B_z^2+g_{\perp}^2B_{\perp}^2}$,
$ \sigma = \pm 1/2 $ is the eigenvalue of $S_z$, $ \alpha = \pm 1 $ is
the  the eigenstate chirality in the limit of vanishing electric
field: $ | \lambda_{\sigma}^{\alpha} \rangle_{\mathcal {E}=0} = ( -
\alpha )^{\sigma -1/2} | \alpha , \sigma \rangle $.  In the presence
of electric field, the eigenstates read:
\begin{eqnarray}\label{epreq2}
| \lambda^{\alpha}_{\sigma} \rangle & = &  \{ 2\sigma [ \Delta\sub{SO}
  + \alpha (\mathcal{E}^2 + \Delta\sub{SO}^2)^{1/2}] |+1,\sigma\rangle
\nonumber\\  & + & \mathcal{E} e^{-i\theta} |-1,\sigma\rangle \} /
D^{\alpha} ,
\end{eqnarray}
where $ D^{\alpha} = \{ \mathcal{E}^2 + [ \Delta\sub{SO} +\alpha
  (\mathcal{E}^2 + \Delta\sub{SO}^2)^{1/2} ]^2 \}^{1/2} $.  

Electron spin resonance induces transitions between such  eigenstates.
The transition amplitudes are given by the absolute values of matrix
elements of $x$-component of the total spin, taken between the states
that the transition connects,
\begin{eqnarray}
\label{EPR1}
\langle \lambda_{-1/2}^{\alpha} | S_x | \lambda_{+1/2}^{-\alpha}
\rangle  & = & - \mathcal{E}^2  / D^{+ 1} D^{- 1} \\ 
\label{EPR2}
\langle \lambda_{-1/2}^{\alpha} | S_x | \lambda_{+1/2}^{\alpha}
\rangle  & = & \frac{\Delta\sub{SO} [ \Delta\sub{SO} +\alpha (\mathcal{E}^2
    + \Delta\sub{SO}^2)^{1/2} ]} {( D^{\alpha} )^2} .
\end{eqnarray}
\begin{figure}
\includegraphics[width=\columnwidth]{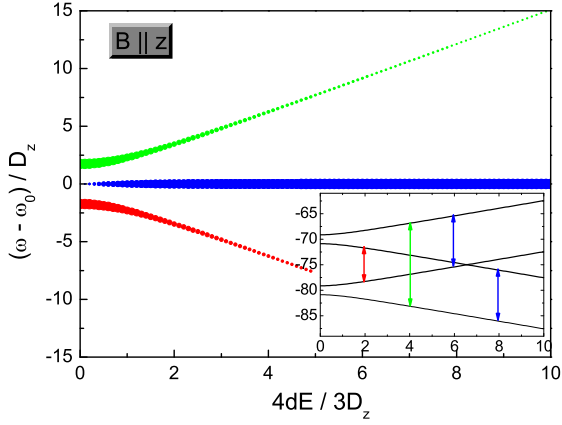}
\caption{\label{fEPR1} 
(color online) Energy ($ \omega $) of the ESR transitions in a triangle of $s=1/2$ spins as a 
function of the applied electric field $\vctr{E}$ that lies in the
molecule's plane, so that $d|\vctr{E}|=dE=\mathcal{E}$.  The magnetic
field is $ {\bf B} \parallel \hat{\bf z} $ and $ \omega_0 = g\mu_B B
$, see Eqs. (\ref{EPR3}) and (\ref{EPR4}).
The diameter of the circles is proportional to the transition
amplitudes $ | \langle \alpha | S_x | 
\alpha' \rangle | $, Eqs. (\ref{EPR1}) and (\ref{EPR2}).  Here, $
|\alpha\rangle $ are the eigenstates of $H$ in the lowest energy
$S=1/2$ multiplet.  Inset: Eigenvalues (in units of $D_z$) as a
function of $\mathcal{E}=d|\vctr{E}_{\parallel}|$, in units $ 3 D_z / 4 $.}
\end{figure}
The corresponding frequencies are given by:
\begin{eqnarray}
\label{EPR3}
\lambda^{\alpha}_{+1/2} - \lambda^{-\alpha}_{-1/2} & = & \mathcal{B}
\\
\label{EPR4}
\lambda^{\alpha}_{+1/2} - \lambda^{\alpha}_{-1/2} & = & \mathcal{B}
+\alpha (\mathcal{E}^2+\Delta\sub{SO}^2)^{1/2} .
\end{eqnarray}
As an illustrative example, we plot the frequencies and amplitudes of
the ESR  transitions as a function of the electric field
(Fig. \ref{fEPR1}). While for  $ \mathcal{E} = 0 $, these transitions
can only take place between states of equal $C_z$ (red and green
symbols online, transitions with the larger amplitude at low fields,
in the figure and in the inset), the  electric field mixes states of
opposite chirality, thus transferring oscillator  strength to two
further transitions, whose frequencies are independent of
$\mathcal{E}$ (blue symbols online, constant frequency transition in
the figure). In the limit $ dE \gg D_z $, the eigenstates of   the
spin Hamiltonian tend to coincide with those of $ {\bf S}^2_{12} $,
and  ESR transitions take place between states of equal $S_{12}$.
While the eigenstates depend on the in-plane orientation of the
electric  field, no such dependence is present in the frequencies and
oscillator  strength of the ESR transitions. Besides, these quantities
are independent of  the exchange coupling $J$, and depend on the value
of the applied magnetic  field only through an additive constant
($\omega_0$).

The dependence of the ESR spectrum on the applied electric field is
qualitatively different if the static magnetic field is applied
in-plane  (e.g., ${\bf B} \parallel \hat{\bf x}$ and the oscillating
field oriented  along $\hat{\bf z}$). In this case, the eigenvalues of
$ H\sub{eff} $ are:
\begin{eqnarray}
\mu^\alpha_\sigma & = &  \alpha\sigma [ \Delta\sub{SO}^2 +
  (\mathcal{E}+\alpha\mathcal{B})^2 ]^{1/2} ,
\end{eqnarray}
where $ \sigma = \pm 1/2 $ is the value of $ \langle S_x \rangle $ in
the limit of large magnetic field ($ \mathcal{B} \gg \mathcal{E},
\Delta\sub{SO} $) and $\alpha = \pm 1$.  The corresponding eigenstates
read:
\begin{eqnarray}
| \mu^\alpha_\sigma \rangle & = & \{ e^{i\theta}  ( \Delta\sub{SO} +
\mu^\alpha_\sigma  )  [ |\!+\!1,\!+\!1/2\rangle -
  |\!-\!1,\!-\!1/2\rangle ] \nonumber\\ & + & ( \mathcal{B} +\alpha
\mathcal{E} )  [ |\!+\!1,\!-\!1/2\rangle - |\!-\!1,\!+\!1/2\rangle ]
\} / D^\alpha_\sigma ,
\end{eqnarray}
where
\begin{eqnarray}
 D^\alpha_\sigma & = &  \sqrt{2} [( \Delta\sub{SO} + \mu^\alpha_\sigma )^2
   + ( \mathcal{B} + \alpha \mathcal{E} )^2]^{1/2} .
\end{eqnarray}
The expectation values of the total spin along the magnetic field for
each of the above eigenstates are given by the following expressions
\begin{eqnarray}
\langle \mu^\alpha_\sigma | S_x | \mu^\alpha_\sigma \rangle & = &  2 [(
  \Delta\sub{SO} + \mu^\alpha_\sigma )( \mathcal{B} - \alpha\mathcal{E} )]
/  (D^\alpha_\sigma )^2 ,
\end{eqnarray}
which are independent of the in-plane direction of the electric  field.  
The ESR transitions between such eigenstates induced by a magnetic field 
that  oscillates along the $z$ direction are given by the expressions:
\begin{eqnarray}
\langle \mu^\alpha_\sigma | S_z | \mu^{-\alpha}_{\sigma'} \rangle  & = &
\frac{(\Delta\sub{SO}+\mu^\alpha_\sigma)(\Delta\sub{SO}+\mu^{-\alpha}_{\sigma'})+(\mathcal{E}^2-\mathcal{B}^2)}{D^\alpha
  D^{-\alpha}},  \nonumber\\  \langle \mu^\alpha_\sigma | S_z |
\mu^\alpha_{\sigma'} \rangle & = & 0.
\end{eqnarray}
Therefore, the application of the electric field shifts the energy of 
the transitions between states of opposite $ \alpha $, thus removing 
their degeneracy; however, unlike the case $ {\bf B} \parallel \hat{\bf z}$, 
it does not increase the number of allowed transitions. 

\begin{figure}
  \includegraphics[width=\columnwidth]{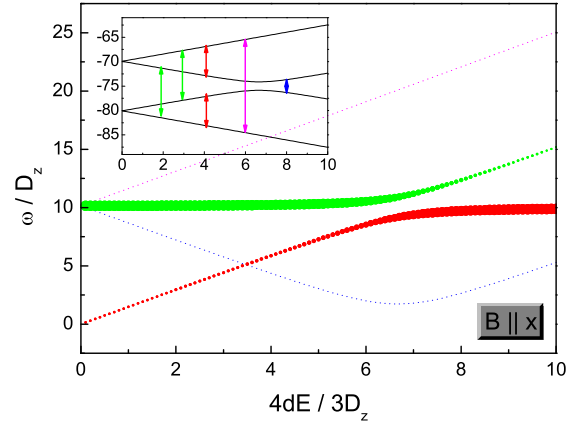}
\caption{\label{fEPR2} Energy ($ \omega $) of the ESR transitions in a triangle of $s=1/2$ spins as a 
function of the applied in-plane electric field $\vctr{E}$, so that
$d|\vctr{E}|=dE=\mathcal{E}$, and in the presence of the in-plane
magnetic field $ {\bf B} \parallel \hat{\bf x} $.
The diameter of the circles is proportional to $ | \langle \alpha | S_z | 
\alpha' \rangle | $, Eqs. (\ref{EPR3}) and (\ref{EPR4}).  The states $
|\alpha\rangle $ are the eigenstates of $H$ in the lowest $S=1/2$ multiplet. 
Inset: Eigenvalues (in units of $D_z$) as a function of
$d|\vctr{E}_{\parallel}|=\mathcal{E}$, in units $ 3 D_z / 4 $.}
\end{figure}

In the case of tilted magnetic fields, the dependence of the ESR
spectrum on  the applied electric field presents qualitatively
different features (Fig. \ref{fEPR2}). In particular, the spectrum is
dominated by two pairs of degenerate transitions that anticross as a
function of the electric field. Away from the anticrossing, the
transitions with the largest oscillator strength display frequency
dependence on the electric field.

\subsubsection{Pentagons of $s=1/2$ spins}

\begin{figure}
\includegraphics[width=\columnwidth]{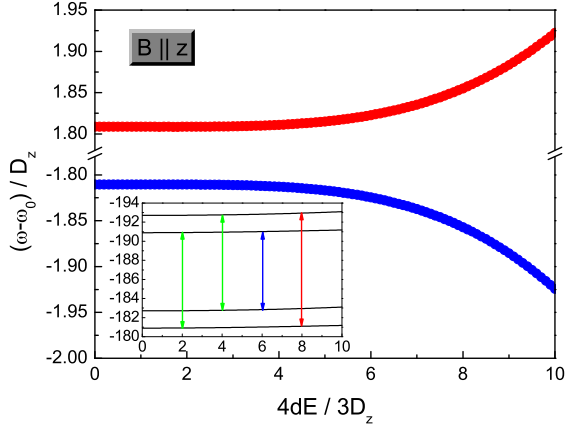}
\caption{\label{fEPR3} 
Energy ($ \omega $) of the ESR transitions in a pentagon of $s=1/2$ spins as a 
function of the electric field applied in the molecule's plane
$d|\vctr{E}|=dE=\mathcal{E}$.  The Zeeman splitting, $ \omega_0 =
g\mu_B B$ is set by the magnetic field 
$ {\bf B} \parallel \hat{\bf z} $, orthogonal to the molecule's plane.
The considered transitions are those between eigenstates ($ |\alpha\rangle $) 
belonging to the $ S = 1/2 $ multiplet of the spin Hamiltonian (figure inset). 
Unlike the case of the spin triangle, these are coupled to each other by the
electric field via eigenstates belonging to other multiplets, 
and therefore depends also on the exchange constant $J$ 
(here $J/\Delta\sub{SO}=100$).
The diameter of the circles is proportional to 
$ | \langle \alpha | S_x | \alpha' \rangle | $,
and therefore to the transition amplitude.}
\end{figure}

\begin{figure}
\includegraphics[width=\columnwidth]{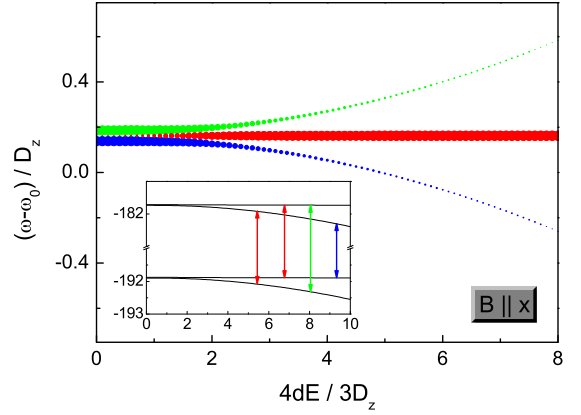}
\caption{\label{fEPR4} 
Energy ($ \omega $) of the ESR transitions in a pentagon of $s=1/2$ spins as a 
function of the applied in-plane electric field $\vctr{E}$, so that
$d|\vctr{E}|=dE={\mathcal{E}}$.  The Zeeman splitting is set by an
in-plane magnetic field $ {\bf B} \parallel \hat{\bf x} $, and $ \omega_0 = g\mu_B B$.
The considered transitions are those between eigenstates ($ |\alpha\rangle $) 
belonging to the $ S = 1/2 $ multiplet of the spin Hamiltonian (figure inset). 
Unlike the case of the spin triangle, these are coupled to each other by the
electric field via eigenstates belonging to other multiplets, 
and therefore depends also on the exchange constant $J$ 
(here $J/\Delta\sub{SO}=100$).
The diameter of the circles is proportional to 
$ | \langle \alpha | S_z | \alpha' \rangle | $,
and therefore to the transition amplitude.}
\end{figure}

Triangles of $s=3/2$ spins (not shown here) display the same
qualitative  behavior as the one discussed above. In contrast, chains
including an  odd number $N>3$ spins behave differently. This is
mainly due to the fact that the spin-electric coupling $\delta H$ does
not couple directly the  four eigenstates of $ H $ belonging to the
lowest $S=1/2$ multiplet: such coupling only takes place through
mixing with the higher $S=1/2$ multiplet.  As a consequence, the
effects of the spin-electric coupling tend to be  weaker as compared
to the case of the triangle; besides, unlike the above case of the
spin triangle, they depend on the exchange coupling $J$.  Illustrative
numerical results are shown in Figs. \ref{fEPR3} and  \ref{fEPR4} for
the cases of a perpendicular and in-plane magnetic field,
respectively. In the former case, both the frequencies and amplitude
of  the ESR transitions are hardly affected by the electric field, in
the  same range of physical parameters considered in Fig. \ref{fEPR1}.
In the case of an in-plane magnetic field, instead, a relatively small
shift in the transition energies is accompanied by a strong transfer
of the oscillator strength, for values of the spin-electric coupling
exceeding the Dzyalozhinsky-Moriya coupling constant.

\subsection{Nuclear magnetic resonance}

The spin-electric Hamiltonian $ \delta H_0 $ modifies non uniformly
the super-exchange  couplings between neighboring spins. This might
not affect the projection of the total  spin (as in the case $ {\bf B}
\parallel \hat{\bf z} $, see above), but it generally affects the
moment distribution within the spin chain. Such effect can be
investigated through  experimental techniques that act as local probes
in molecular nanomagnets, such as  nuclear magnetic resonance (NMR)
\cite{MFK+06} or x-ray absorption \cite{GLM+09}.  In NMR, the
expectation value of a given spin within the cluster can be inferred
through the frequency shift induced on the transitions of the
corresponding nucleus.  The shift in the nuclear resonance frequency
for the nucleus of the $i$-th  magnetic  ion is $ \Delta \nu = \gamma
A \langle s_{z,i} \rangle $, where $A$ is the contact hyperfine
interaction constant at the nuclear site. The constant of
proportionality  $ A $ depends on the spin density at the position of
the nucleus, and can be extracted  from the experiment by considering
the polarized ground state $ M = S $ at high  magnetic fields
\cite{MFK+06}.  As in the case of ESR, the dependence of the NMR
spectra on the applied  electric field  qualitatively depends on the
orientation of the static magnetic field ${\bf B}$ with  respect to
the molecule. Unlike the case of ESR, however, it also depends on the
in-plane orientation of the electric field, i.e. on the way in which
the $ {\bf E} $ breaks the symmetry of the molecule. 

\subsubsection{Spin triangles}

Let us start by considering a spin $ s = 1/2 $ triangle, with a
magnetic field applied  perpendicular to the molecule plane ($ {\bf B}
\parallel \hat{\bf z} $).  In this case, the distribution of the spin
projection along $z$ is given by the following expression:
\begin{equation}
\label{eq:sznmr}
\langle \lambda_{\sigma}^{\alpha} | s_{i,z} |
\lambda_{\sigma}^{\alpha} \rangle  =  \sigma / 3 +
f_{\sigma}^{\alpha} (\mathcal{E}) \cos [ \theta + \pi (5/3-i) ] ,
\end{equation}
where
\begin{equation}
f_{\sigma}^{\alpha} (\mathcal{E}) \equiv
\frac{4\sigma\mathcal{E}[\Delta\sub{SO}+\alpha
    (\Delta\sub{SO}^2+\mathcal{E}^2)^{1/2}]}{3 (D^{\alpha})^2 } .
\end{equation}
Here, the expressions of the eigenstates $ | \lambda^\alpha_\sigma \rangle $ 
and of $ D^\alpha $ are given in Subsection \ref{ssEPR}.
For $E=0$, the three spins are equivalent and  $ \langle
\lambda^{\alpha}_{\pm 1/2} | s_{i,z} | \lambda^{\alpha}_{\pm 1/2}
\rangle  = \pm 1/6 $.  
\begin{figure}
\includegraphics[width=\columnwidth]{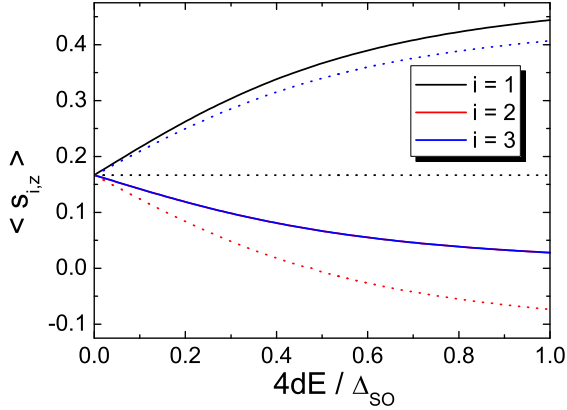}
\caption{\label{figNMR1} (Color online) Expectation values of the 
 $z$-component of $s=1/2$ spins in a triangular molecule as a function
  of applied electric field.   The magnetic field is perpendicular to the ring plane ($ {\bf B} \parallel \hat{\bf z}$); the electric field is parallel and 
perpendicular to ${\bf r}_{12}$ in the upper and lower panel,
respectively. In the
  electric field along one of the bonds (lower panel), the spins that
  lie on that bond have the same out-of-plane projections.  The
  shadings (colors online) denote the different spins.}
\end{figure}
\begin{figure}
\includegraphics[width=\columnwidth]{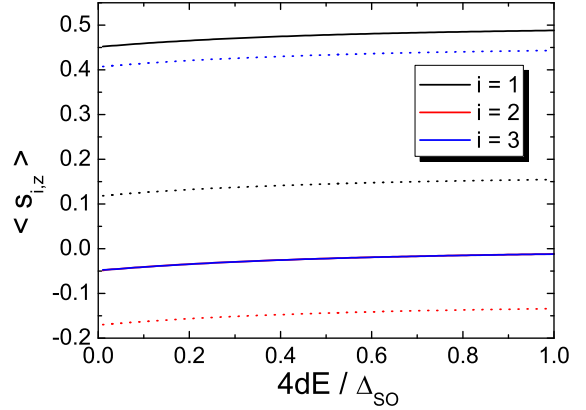}
\caption{\label{figNMR2} (Color online) Expectation values of the
  $z$-component of $s=3/2$ spins in a triangular molecule as a
  function of applied electric field.   The magnetic field is
  perpendicular to the ring plane ($ {\bf B} \parallel \hat{\bf z}$);
  the electric field is parallel and  perpendicular to ${\bf r}_{12}$
  in the upper and lower panel, respectively.  The shadings (colors
  online) denote the different spins.}
\end{figure}
If the electric field is finite and oriented along one of the triangle
sides (e.g., ${\bf E} \parallel  {\bf r}_{12}$, corresponding to
$\theta  = 0 $), then expectation values along $z$ of spins 1 and 2
undergo  opposite shifts, whereas that of spin 3 is left unchanged:  $
\Delta_{\bf E} \langle s_{1,z} \rangle =   - \Delta_{\bf E} \langle
s_{2,z} \rangle $, where   $ \Delta_{\bf E} \langle s_{i,z} \rangle
\equiv  \langle s_{i,z} \rangle_{\bf E}  -  \langle s_{i,z}
\rangle_{{\bf E}=0} $.  This is shown in Fig. \ref{figNMR1} for the
ground state of the spin  Hamiltonian, but the above relations hold
for any of the four  eigenstates $ |\lambda^\alpha_\sigma \rangle $
belonging to the $S=1/2$ quadruplet.   If the NMR frequency shifts $
\Delta\nu_i $ are larger than the  corresponding line widths, the
single line at ${\bf E}=0$ splits into  three equispaced  lines, with
intensity ratios 1:1:1. If, instead, the  electric field is applied
along a symmetry plane of the triangle (e.g.,  $ {\bf E} \perp {\bf
  r}_{12}$, corresponding to $ \theta = \pi / 2 $),  spins 1 and 2
remain equivalent and their magnetic moments display  the same
electric field dependence, while the shift of the third one is
opposite in sign and twice as large  in absolute value:   $
\Delta_{\bf E} \langle s_{1,z} \rangle = \Delta_{\bf E} \langle
s_{2,z} \rangle =    - \Delta_{\bf E} \langle s_{3,z} \rangle / 3 $.
The intensity ratios of the two NMR lines are, correspondingly, 1:2.
The expectation values for the remaining eigenstates can be derived by
the following  equations: $ \Delta_{\bf E}  \langle
\lambda^{\alpha}_{-1/2} | s_{i,z} | \lambda^{\alpha}_{-1/2} \rangle  =
- \Delta_{\bf E}  \langle \lambda^{\alpha}_{+1/2} | s_{i,z} |
\lambda^{\alpha}_{+1/2} \rangle $ and $ \langle \lambda^{ 1}_{\sigma}
| s_{i,z} | \lambda^{ 1}_{\sigma} \rangle   =  - \Delta_{\bf E}
\langle \lambda^{-1}_{\sigma} | s_{i,z} | \lambda^{-1}_{\sigma}
\rangle $.  Therefore, at finite temperature, the shifts in the
expectation values of the  three spins are given by:
\begin{equation}
\frac{ \Delta_{\bf E} \langle s_{i,z} \rangle} {\Delta_{\bf E} \langle
  \lambda^{+1}_{-1/2} | s_{i,z} | \lambda^{+1}_{-1/2} \rangle}  =
\frac{ \sum_{\alpha} \alpha \cosh \left( \frac{\lambda^{\alpha}_{-1/2}
  }{ k_B T } \right) } { \sum_{\alpha}        \cosh \left(
  \frac{\lambda^{\alpha}_{-1/2} }{ k_B T } \right) } .
\end{equation}

If the field is oriented along the molecule plane ($ {\bf B} \parallel
\hat{\bf x}$), the expectation value of the three spins corresponding
to  each of the eigenstates are given by the following expressions:
\begin{eqnarray}
\langle \mu_{\sigma}^{\alpha} | s_{i,x} | \mu_{\sigma}^{\alpha} \rangle
=  g_{\sigma}^{\alpha} (\mathcal{E}) + (1/3) \cos ( \theta - 2 i \pi /
3 ) ,
\end{eqnarray}
where
\begin{equation}
g_{\sigma}^{\alpha} (\mathcal{E}) \equiv  \frac{2}{3} \frac{(
  \Delta\sub{SO} + \mu_{\sigma}^{\alpha} ) ( \mathcal{B} +\alpha
  \mathcal{E} )}{(D^\alpha)^2} .
\end{equation}
If the magnetic field is parallel to the triangle plane, the in-plane
electric field can modify the total spin expectation value along
${\bf B}$. The changes that $ {\bf E} $ induces in the magnetization
distribution  within the triangle at zero temperature are less varied
than in the previous case (Fig. \ref{figNMR1}). In fact, the magnitude
of  the $ \Delta_{\bf E} \langle s_{i,z} \rangle $ is much smaller,
and all  the spins undergo shifts of equal sign and slope. The NMR
line, which  is slitted into three lines already for ${\bf E}=0$, is
rigidly by the  applied electric field.

If the triangle is formed by half-integer spins $ s > 1/2 $, an
analogous  dependence of the expectation values $ \langle s_{i,z}
\rangle $ on the  electric field is found. As an illustrative example,
we report in Fig.  \ref{figNMR2} the case of $s = 3/2 $. 

\subsubsection{Pentagon of $s=1/2$ spins}

\begin{figure}
\includegraphics[width=\columnwidth]{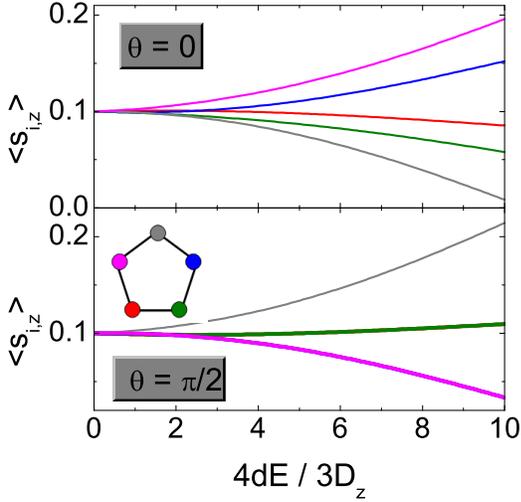}
\caption{\label{figNMR3} (Color online) Expectation values of the 
 $z$-component of $s=1/2$ spins in a pentagon as a function of applied
  electric field.   The magnetic field is perpendicular to the ring
  plane ($ {\bf B} \parallel \hat{\bf z}$); the electric field is
  parallel ($\theta=0$) and 
perpendicular ($\theta=\pi/2$) to ${\bf r}_{12}$ in the upper and lower panel,
respectively.  The
  shadings (colors online) denote the different spins.}
\end{figure}

Spin chains consisting of an odd number of half-integer spins present 
analogous behaviors, but also meaningful differences with respect to the
case of the spin triangle. In particular, the spin-electric Hamiltonian
$ \delta H_0 $ does not couple states belonging to the lowest $ S=1/2 $ 
quadruplet directly (i.e., matrix elements $ \langle i | \delta H_0 | j
\rangle = 0 $ for $i,j \le 4$); these couplings are mediated by states
belonging to higher $ S = 1/2 $ multiplets, that are higher in energy by
a quantity $ \sim J $. Therefore, the effect of the electric field tends 
to be significantly smaller than in the case of a triangle with equal 
$ D_z $ and $ \mathcal{E} $ (see Fig. \ref{figNMR3}), and depends also 
on the exchange coupling $J$.

%

\subsection{Magnetization, Polarization, and Susceptibilities}

The spin-electric coupling shifts the energy eigenvalues of the
nanomagnet, thus affecting thermodynamic quantities, such as
magnetization, polarization and susceptibilities.  In the following,
we compute these quantities in  the case of the $s=1/2$ spin triangle
as  a function of the applied magnetic and electric fields.  Under the
realistic assumption that the exchange splitting $J$ is the largest
energy scale in the spin Hamiltonian, and being mainly interested in
the low-temperature limit, we restrict ourselves to the $S=1/2$
quadruplet, and use the effective Hamiltonian $H\sub{eff}$ in
Eq. (\ref{eff_Ham_three_half}).   

The eigenenergies of the lowest $S=1/2$ sector in the presence of
electric and magnetic fields are
\begin{eqnarray}
E_{\alpha,\gamma}&=&\alpha\gamma\sqrt{\mathcal{B}^2+\Delta\sub{SO}^2+\mathcal{E}^2+2\gamma
  E_0^2},
\label{eigenstates_low_energy}
\end{eqnarray}  
with
$\mathcal{B}=\mu_B\sqrt{g_{\parallel}^2H_{\parallel}^2+g_{\perp}^2H_{\perp}^2}$,
$E_0=[(\mathcal{B}_z\Delta\sub{SO})^2+(\mathcal{B}\mathcal{E})^2)]^{1/4}$,
and $\mathcal{B}_z=\mu_Bg_{\parallel}H_{\parallel}$. Note that these
energies are the generalization of the ones in the previous section,
which were valid for in-plane magnetic field only.  The partition
function for N identical and non-interacting molecules is $Z=Z_1^N$,
with $Z_1=\sum_{\alpha,\gamma}\exp{(-\beta E_{\alpha,\gamma})}$ being
the partition function  for one molecule, and
$\beta=1/(k\sub{B}T)$. The free energy  reads 
\begin{equation}
F\equiv -1/\beta\ln{Z}=-Nk\sub{B}T\ln{\left[2\sum_{\gamma}\cosh{(\beta
      E_{\gamma})}\right]},
\label{free_energy}
\end{equation}
with $E_{\gamma}\equiv E_{1/2,\gamma}$.  From this, we can derive
different thermodynamic quantities like the  magnetization
$M_i=-\partial F/\partial H_{i}$, the electric polarization
$P_i=-\partial F/\partial E_i$, the heat capacity
$C=-\partial/\partial T(\partial \ln{(Z)}/\partial \beta)$, and the
corresponding susceptibilities: $\chi_{E_iE_j}=\partial P_i/\partial
E_j=\partial^2F/\partial E_i\partial E_j$ - the {\it electric}
susceptibility,  $\chi_{H_iH_j}=\partial M_i/\partial
H_j=\partial^2F/\partial H_i\partial H_j$ - the {\it spin}
susceptibility, and  $\chi_{E_iH_j}=\partial P_i/\partial
M_j=\partial^2F/\partial E_i\partial H_j$ - the {\it spin-electric}
susceptibility.  For the electric polarization components $P_i$ we
get
\begin{eqnarray}
P_i&=&\frac{Nd\mathcal{E}_i}{4\sum_{\gamma=\pm1}\cosh{(\beta
    E_{\gamma})}}\sum_{\gamma=\pm1}\frac{\sinh{(\beta
    E_{\gamma})}}{E_{\gamma}}\nonumber\\ &\times&\left(1+\gamma\frac{\mathcal{B}^2}{E_0^2}\right)(1-\delta_{i,z}),
\label{eq:polarization}
\end{eqnarray}
while for the magnetization components $M_i$ we get
\begin{eqnarray}
M_i&=&\frac{Ng_i\mu_B\mathcal{B}_i}{2\sum_{\gamma=\pm1}\cosh{(\beta
    E_{\gamma})}}\sum_{\gamma=\pm1}\frac{\sinh{(\beta
    E_{\gamma})}}{E_{\gamma}}\nonumber\\ &\times&\left(1+\gamma\frac{\Delta\sub{SO}^2\delta_{i,z}+\mathcal{E}^2}{E_0^2}\right),
\label{eq:magnetization}
\end{eqnarray}
where again $i=x,y$.  Making use of the above expressions, we can
obtain the above defined susceptibilities
\begin{widetext}
\begin{eqnarray}
\chi_{E_iE_j}&=&\frac{P_i}{E_j}\delta_{ij}-\beta
P_iP_j+\frac{Nd^4E_iE_j}{2\sum_{\gamma=\pm1}\cosh{(\beta
    E_{\gamma})}}\nonumber\\ &\times&\bigg[\sum_{\gamma=\pm1}\frac{\beta
    E_{\gamma}\cosh{(\beta E_{\gamma})}-\sinh{(\beta
      E_{\gamma})}}{2E_{\gamma}^3}\left(1+\gamma\frac{\mathcal{B}^2}{E_0^2}\right)^2+\gamma\frac{\mathcal{B}^4}{E_0^6}\frac{\sinh{(\beta
      E_{\gamma})}}{E_{\gamma}}\bigg]=\chi_{E_jE_i} \label{eq:es}\\ \chi_{B_iB_j}&=&\frac{M_i}{B_j}\delta_{ij}-\beta
M_iM_j+\frac{Ng_i^2g_j^2B_iB_j}{2\sum_{\gamma=\pm1}\cosh{(\beta
    E_{\gamma})}}\sum_{\gamma=\pm1}\bigg[\frac{(\Delta\sub{SO}^2\delta_{i,z}+\mathcal{E}^2)(\Delta\sub{SO}^2\delta_{j,z}+\mathcal{E}^2)}{E_0^6}\frac{\sinh{(\beta
      E_{\gamma})}}{E_{\gamma}}\nonumber\\ &+&\frac{\beta
    E_{\gamma}\cosh{(\beta E_{\gamma})}-\sinh{(\beta
      E_{\gamma})}}{2E_{\gamma}^3}\left(1+\gamma\frac{\Delta\sub{SO}^2\delta_{i,z}+\mathcal{E}^2}{E_0^2}\right)\left(1+\gamma\frac{\Delta\sub{SO}^2\delta_{j,z}+\mathcal{E}^2}{E_0^2}\right)\bigg]=\chi_{B_jB_i}
\label{eq:ms}\\ \chi_{B_iE_j}&=&-\beta
M_iP_j+\frac{Ng_i^2d^2B_iE_j}{2\sum_{\gamma=\pm1}\cosh{(\beta
    E_{\gamma})}}\sum_{\gamma=\pm1}\bigg[\gamma\frac{(\Delta\sub{SO}^2\delta_{i,z}+\mathcal{E}^2)(\Delta\sub{SO}^2\delta_{j,z}+\mathcal{E}^2)}{E_0^6}\frac{\sinh{(\beta
      E_{\gamma})}}{E_{\gamma}}\nonumber\\ &+&\frac{\beta
    E_{\gamma}\cosh{(\beta E_{\gamma})}-\sinh{(\beta
      E_{\gamma})}}{2E_{\gamma}^3}\left(1+\gamma\frac{\Delta\sub{SO}^2\delta_{i,z}+\mathcal{E}^2}{E_0^2}\right)\left(1+\gamma\frac{\mathcal{B}^2}{E_0^2}\right)\bigg](1-\delta_{j,z})=\chi_{E_jB_i}.
\label{eq:sels}
\end{eqnarray}
\end{widetext}
The polarization $\vctr{P}$, magnetization $\vctr{M}$, and
susceptibilities $\chi$, Eq. (\ref{eq:polarization}) --
Eq. (\ref{eq:sels}),  all depend on the spin-electric coupling
constant $d$.  In the following, we analyze the details of this
dependence and identify the conditions suitable for extracting the
value of $d$ from the measurable quantities.

\subsubsection{Polarization and magnetization}

\begin{figure}[t]
\begin{center}
\includegraphics[width=6.5cm]{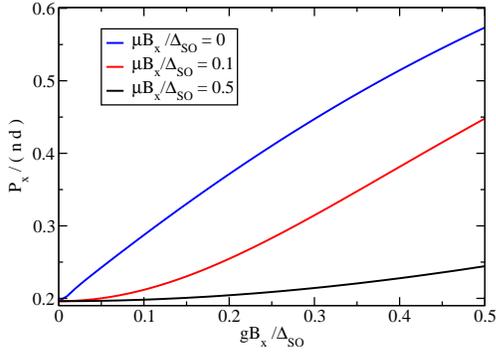}
\vspace{0.2cm}
\caption{\label{fig:pxbx}  Electric polarization $P_x$ ($x$ component) in Eq. (\ref{eq:polarization})  as a
  function of the magnetic field in $x$ direction.  The three lines
  correspond to various values of an additional external electric field in the
  $z$ direction.  The plot is for the temperature $k\sub{B}T =
  0.001\Delta\sub{SO}$, and the electric field $dE_x = 0.1\Delta\sub{SO}$.} 
\end{center}
\end{figure}

\begin{figure}[t]
\begin{center}
\includegraphics[width=6.5cm]{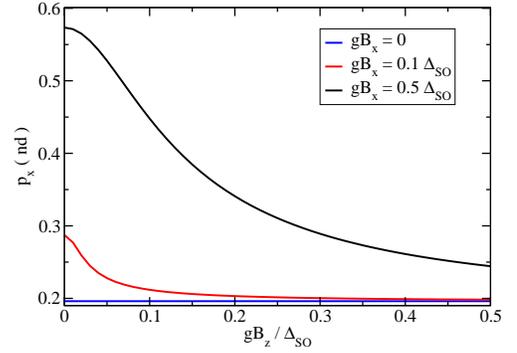}
\vspace{0.2cm}
\caption{\label{fig:pxbz}  Electric polarization $P_x$ ($x$ component) in  Eq. (\ref{eq:polarization}) as a
  function of the magnetic field in $z$ direction.  The three lines
  correspond to various values of the external magnetic field in the
  $x$ direction.  The plot is for the temperature $k\sub{B}T =
  0.001\Delta\sub{SO}$, and the electric field $dE_x = 0.1\Delta\sub{SO}$.} 
\end{center}
\end{figure}

The in-plane polarization of the molecule as a function of the
magnetic field is illustrated in Fig. \ref{fig:pxbx} and
Fig. \ref{fig:pxbz}.  The polarization is a growing function of the
magnetic field strength, and it gets reduced by the normal component
of the field.  

The low-temperature, $k\sub{B}T \ll \Delta\sub{SO}$, thermodynamic
properties of a molecule with spin-electric coupling show a simple
dependence on the strength of external electric and magnetic fields in
the special cases of in-plane and out-of plane magnetic field.  We
focus only on effects in leading orders in electric field under the
realistic assumption that the electric dipole splitting is small
compared to the SO splitting, i.e. $\mathcal{E}\ll\Delta\sub{SO}$.  Also,
we analyze two limiting cases: (i) $k\sub{B}T \ll \mathcal{E}$,
i.e. low-temperature regime, and (ii) $ k\sub{B}T \gg \mathcal{E}$,
i.e. high temperature regime.  However, we assume all temperatures (in
both regimes) to satisfy $k\sub{B}T \ll \Delta\sub{SO}$ so that the
spin-orbit split levels are well resolved. In the first case (i), we
obtain for the polarization 
\begin{eqnarray}
P_i\simeq\left\{
\begin{array}{ll} \displaystyle \frac{nd\mathcal{E}_i\mathcal{B}}{4\mathcal{E}\Delta_{\mathcal{B}}} & \textrm{ for $\mathcal{E}\ll \mathcal{B}$}\vspace{0.2cm}\\
\displaystyle
\frac{nd\Delta\sub{SO}^2\mathcal{E}_i}{4\Delta_{\mathcal{B}}^3} & \textrm{ for $\mathcal{E}\gg \mathcal{B}$}\,
,\end{array}\right.
\end{eqnarray}
while for the second situation (ii) we obtain
\begin{eqnarray}
 P_i&\simeq&\frac{nd\Delta\sub{SO}^2\mathcal{E}_i}{4\Delta_{\mathcal{B}}^3}\left(1+\frac{\mathcal{B}^2}{\Delta\sub{SO}^2}\beta\Delta_{\mathcal{B}}\right),
\end{eqnarray}
with $\Delta_{\mathcal{B}}=\sqrt{\mathcal{B}^2+\Delta\sub{SO}^2}$ and
$n=N/V$ the density of molecules in the crystal. We see that, for low
temperatures, the electric polarization $P_i$ ranges from being
independent of the magnitude of the electric field
($\mathcal{E}\ll\mathcal{B}$), to a linear dependence on the applied
electric field $E$ for large fields
($\mathcal{E}\gg\mathcal{B}$). Also, the polarization is strongly
dependent on the magnetic field (linear in $B$) for low $E$-fields,
thus implying  strong magneto-electric response.

We now switch to the other special case, namely when the external
magnetic field is applied perpendicularly to the spin triangles.  The
electric polarization now reads 
\begin{equation}
P_i=\frac{nd\mathcal{E}_i}{4\Delta_{\mathcal{E}}}\tanh{(\beta\Delta_{\mathcal{E}})}, 
\end{equation}
with $\Delta_{\mathcal{E}}=\sqrt{\Delta\sub{SO}^2+\mathcal{E}^2}$.
The polarization $P_i$ does not depend on the magnetic field $B$, and
there are no spin-electric effects present for this particular case. 

Our results suggest that the spin-electric coupling can be detected by
measuring the polarization of the crystal of triangular single
molecule antiferromagnets that lie in parallel planes in the in-plane
electric and magnetic fields.

\begin{figure}[t]
\begin{center}
\includegraphics[width=6.5cm]{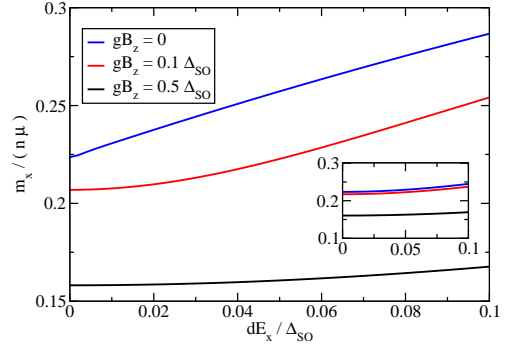}
\vspace{0.2cm}
\caption{
\label{fig:mxex} 
In-plane magnetization $M_x$ in $x$-direction in  Eq. (\ref{eq:magnetization}) as a function of the electric
field $E_x$ in $x$-direction.  The three lines correspond to a fixed value
of an additional magnetic field in the $z$-direction.  The assumed
temperature  is $k\sub{B}T = 0.001\Delta\sub{SO}$, while in the inset it is
at higher temperature $k\sub{B}T = 0.1\Delta\sub{SO}$. } 
\end{center}
\end{figure}

The out-of plane component $M_z$ of the molecule's magnetization is
rather insensitive to the electric fields, since any effect of the
applied in-plane electric field has to compete with the spin-orbit
coupling induced zero-field splitting $\Delta\sub{SO}$.  Since we
expect to find weak coupling to electric field and small coupling
constant $d$, it would require very strong electric field to achieve
the regime $d|\vctr{E}| \sim \Delta\sub{SO}$.  The in-plane components
of magnetization $M_x$, $M_y$, on the other hand show clear dependence
on electric fields, Fig. \ref{fig:mxex}.  At low magnetic fields the
in-plane component of polarization appears and grows with the strength
of in-plane electric fields.  However, the electric field dependence
becomes less pronounced when an additional magnetic field is applied
normal to the triangle's plane.

In the dependence of the magnetization on electric fields, and for the
case of an in-plane magnetic field, we find the same two main regimes
as in the study of the polarization: $\mathcal{E}\gg k_BT$ (i) and
$\mathcal{E}\ll k_BT$ (ii). In the first case (i) we obtain
\begin{eqnarray}
M_i\simeq\frac{ng_i\mu_B\mathcal{B}_i}{4\Delta_{\mathcal{B}}}\left(1+\frac{\mathcal{E}\Delta\sub{SO}^2}{\mathcal{B}\Delta_{\mathcal{B}}^2}\right),
\end{eqnarray}
while for the second case (ii) we get
\begin{eqnarray}
M_i&=&\frac{ng_{\perp}\mu_B\mathcal{B}_i}{4\Delta_{\mathcal{B}}}\bigg[1-\frac{3\mathcal{E}^2\Delta\sub{SO}^2}{2\Delta_{\mathcal{B}}^4}\left(1-\frac{\beta\Delta_{\mathcal{B}}}{3}\right)\bigg].
\end{eqnarray}
The magnetization shows a strong dependence on the electric field $E$,
especially for $\mathcal{E}\gg\mathcal{B}$ where this is linear in
$E$-field. For low electric fields, however, the magnetization shows
only a weak dependence on the electric field, both at low and high
temperatures.

For the magnetization (along $z$) in the presence of a perpendicular
(also along $z$) magnetic field we obtain
\begin{eqnarray}
M_z=\frac{ng_z\mu_B}{4}\tanh{(\beta \mathcal{B})},
\end{eqnarray}
which is manifestly independent of the spin-electric coupling constant $d$.

\subsubsection{Susceptibilities}    %

\begin{figure}[t]
\begin{center}
\includegraphics[width=6.5cm]{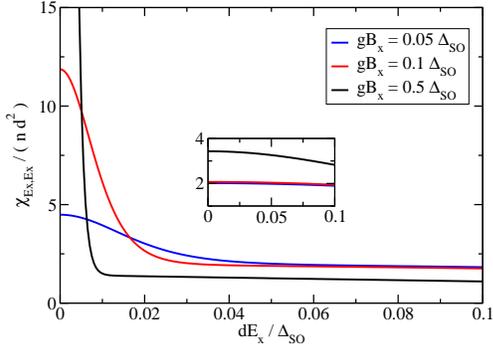}
\vspace{0.2cm}
\caption{\label{fig:esxxex}  Electric susceptibility ($xx$ component),
  Eq. (\ref{eq:es}), 
  as a function of the electric field in $x$ direction.  The three
  lines correspond to various values of the external magnetic field in
  the $x$ direction.  The plot is for the temperature $k\sub{B}T =
  0.001\Delta\sub{SO}$.  In the inset, the same quantity is plotted at
  a higher temperature, $k\sub{B}T = 0.1\Delta\sub{SO}$.  }
\end{center}
\end{figure}

\begin{figure}[t]
\begin{center}
\includegraphics[width=6.5cm]{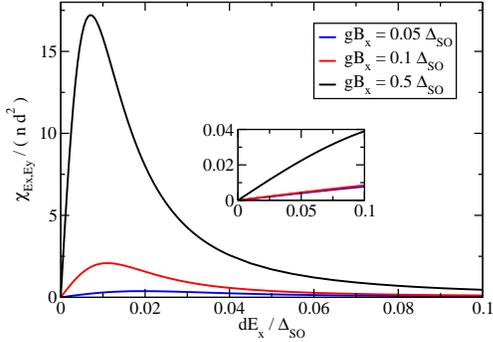}
\vspace{0.2cm}
\caption{\label{fig:esxyex}  Electric susceptibility ($xy$ component), Eq. (\ref{eq:es}),
  as a function of the electric field in $x$ direction.  The three
  lines correspond to various values of the external magnetic field in
  the $x$ direction.  The plot is for the temperature $k\sub{B}T =
  0.001\Delta\sub{SO}$.  In the inset the same quantity is plotted at a higher temperature $k\sub{B}T=0.1\Delta\sub{SO}$. } 
\end{center}
\end{figure}

The effects of spin-electric coupling on the polarization of a
molecule show up in the electric susceptibility and the spin-electric
susceptibility.  In Fig. \ref{fig:esxxex} and Fig. \ref{fig:esxyex},
we plot the $xx$ and $xy$ component of the electric susceptibility
tensor as a function of electric field for various strengths and
orientations of an additional magnetic field.  Both susceptibilities
show a high peak in the region of weak electric fields that is
suppressed by  in-plane magnetic fields.  The peaks are pronounced at
low temperatures, and vanish as the temperature exceeds the splitting
of the two lowest-energy levels, $k\sub{B}T \gg d
|\vctr{E}|_{\parallel}$.  

In the case of in-plane magnetic field, and weak coupling to the
electric field $d|\vctr{E}| \ll \Delta\sub{SO}$, we can calculate the
electric $\chi_{E_iE_j}$ and spin-electric $\chi_{E_iH_j}$
susceptibilities in the two limiting cases (i) and (ii) defined above,
with $i=x,y$.  For the electric susceptibility we obtain:
\begin{eqnarray}
\chi_{E_iE_j}\simeq\left\{\begin{array}{ll} \displaystyle
\frac{nd^2\mathcal{B}(\mathcal{E}^2\delta_{ij}-\mathcal{E}_i\mathcal{E}_j)}{4\mathcal{E}^3\Delta_{\mathcal{B}}}
& \textrm{ for $\mathcal{E}\ll \mathcal{B}$}\vspace{0.2cm}\\ \displaystyle
\frac{nd^2\Delta\sub{SO}^2\delta_{ij}}{4\Delta_{\mathcal{B}}^3} &
\textrm{ for $\mathcal{E}\gg \mathcal{B}$}
\end{array}\right.
\end{eqnarray}
in the first case (i), and 
\begin{eqnarray} 
\chi_{E_iE_j}&\simeq&\frac{nd\Delta\sub{SO}^2\delta_{ij}}{4\Delta_{\mathcal{B}}^3}\left(1+\frac{\mathcal{B}^2}{\Delta\sub{SO}^2}\beta\Delta_{\mathcal{B}}\right).
\end{eqnarray}
in the second case (ii). We see that for low $E$-fields, the electric
susceptibility $\chi_{E_iE_j}$ depends strongly on the applied
electric field, and even vanishes if the field is applied, say, along
$x$ or $y$ directions. For large $E$-fields instead, the electric
susceptibility becomes independent of the electric field itself and,
for low magnetic fields (i.e., for $\mathcal{B}\ll\Delta\sub{SO}$)
this reduces to a constant value $\chi_{E_iE_j}=\delta_{ij}nd^2/4$. At
finite (large) temperatures the electric susceptibility is still
independent of the electric field, but it is  enhanced by thermal
effects $\sim 1/T$.  

For the electric susceptibilities $\chi_{E_iE_j}$ in perpendicular
magnetic field, we obtain
\begin{eqnarray}
\chi_{E_iE_j}&=&\frac{nd^2}{4\Delta_{\mathcal{E}}}\left(\delta_{ij}-\frac{\mathcal{E}_i\mathcal{E}_j}{\Delta_{\mathcal{E}}}\right),
\end{eqnarray}
where we assumed $\Delta\sub{SO}\gg k_BT$, as in the previous
Section. As expected, there is no dependence of $\chi_{E_iE_j}$ on the
$B$-field, and for vanishing electric field the electric
susceptibility reduces to a constant
$\chi_{E_iE_j}=nd^2/4\Delta\sub{SO}$.  

The quantity of most interest in the present spin system is the
spin-electric susceptibility $\chi_{E_iB_j}$, i.e. the magnetic
response (electric response) in electric fields (magnetic fields).
The nonzero spin-electric susceptibility allows for the electric
control of magnetization and magnetic control of polarization in the
crystals of triangular MNs, even in the case when the coupling between
the molecules is negligible.  In addition, $\chi_{E_iB_j}$ is nonzero
only in the presence of spin-electric coupling, i.e. when $d \neq 0$.

\begin{figure}[t]
\begin{center}
\includegraphics[width=6.5cm]{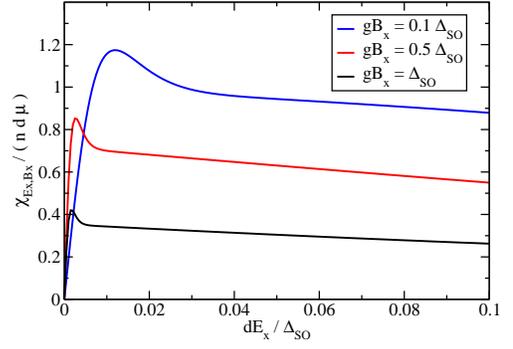}
\vspace{0.2cm}
\caption{\label{fig:lmexxex} Linear magnetoelectric tensor ($xx$
  component) in  Eq. (\ref{eq:sels}) as a function of the electric field in $x$ direction.
  The three lines correspond to various values of the external
  magnetic field in the $x$ direction.  The plot is for the
  temperature $k\sub{B}T = 0.001\Delta\sub{SO}$.  } 
\end{center}
\end{figure}

\begin{figure}[t]
\begin{center}
\includegraphics[width=6.5cm]{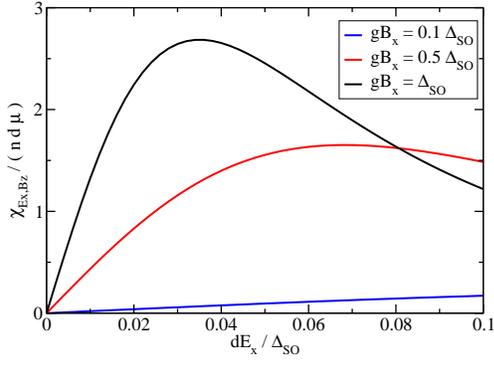}
\vspace{0.2cm}
\caption{\label{fig:lmexzex} Linear magnetoelectric tensor ($xz$
  component) in  Eq. (\ref{eq:sels}) as a function of the electric field in $x$ direction.
  The three lines correspond to various values of the external
  electric field in the $x$ direction.  The plot is for the
  temperature $k\sub{B}T = 0.001\Delta\sub{SO}$.  } 
\end{center}
\end{figure}

The spin-electric susceptibility shows a characteristic peak in weak
electric fields which vanishes in an external magnetic field, see
Figs.~\ref{fig:lmexxex} and \ref{fig:lmexzex}.    
The peak in the diagonal $xx$-component, $\chi_{E_xE_x}$, moves towards
the higher electric fields and broadens as the magnetic field $B_x$
increases.  The peak in the  off-diagonal component $\chi_{E_xB_z}$, on the
other hand, shifts towards the lower electric fields, and narrows as
the in-plane magnetic field increases.  Both peaks disappear at high
temperatures, $k\sub{B} T \gg \Delta\sub{SO}$.
\begin{figure}[t]
\begin{center}
\includegraphics[width=6.5cm]{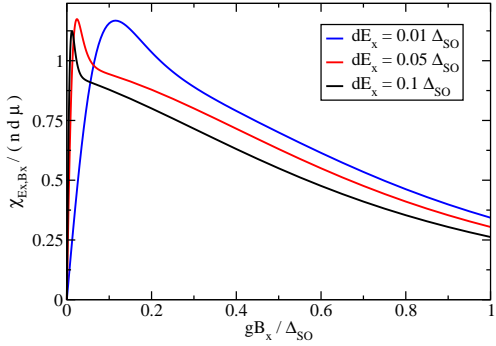}
\vspace{0.2cm}
\caption{\label{fig:lmexxbx} Linear magnetoelectric tensor ($xx$
  component) in  Eq. (\ref{eq:sels}) as a function of the magnetic field in $x$ direction.
  The three lines correspond to various values of the external
  electric field in the $x$ direction.  The plot is for the
  temperature $k\sub{B}T = 0.001\Delta\sub{SO}$.  } 
\end{center}
\end{figure}  

For in-plane magnetic fields and weak spin-electric coupling 
the spin-electric susceptibility $\chi_{E_iB_j}$ is
\begin{eqnarray}
\chi_{E_iB_j}\simeq
\frac{ndg_j\mu_B\mathcal{E}_i\mathcal{B}_j\Delta\sub{SO}^2}{4\mathcal{E}\mathcal{B}\Delta_{\mathcal{B}}^3}
\end{eqnarray}
for the low temperature case (i), while for the second case (ii) we
obtain
\begin{eqnarray}
\chi_{E_iB_j}&\simeq&-\frac{3n\Delta\sub{SO}^2dg_j\mu_B\mathcal{E}_i\mathcal{B}_j}{4\Delta_{\mathcal{B}}^5}\left(1-\frac{\beta\Delta_{\mathcal{B}}}{3}\right).
\end{eqnarray}
By inspecting the above expression, we can infer that for low
temperatures and low $E$-fields the spin-electric susceptibility shows
no dependence on the absolute value of the electric field $E$ and only
a weak dependence on the applied magnetic field $B$. Moreover, when
both fields are applied along one special direction, say, along $x$,
and assuming also $\mathcal{B}\ll\Delta\sub{SO}$, the spin-electric
susceptibility becomes $\chi_{E_xB_x}=ndg_i\mu_B/4\Delta\sub{SO}$,
i.e. it reaches a constant value.  The finite temperature expression
shows  that the spin-electric response is reduced, as opposed to the
electric response where temperature increases the response. Thus, for
strong spin-electric response one should probe the spin system at low
temperatures ($k_BT\ll\Delta\sub{SO}$). 

\begin{figure}[t]
\begin{center}
\includegraphics[width=6.5cm]{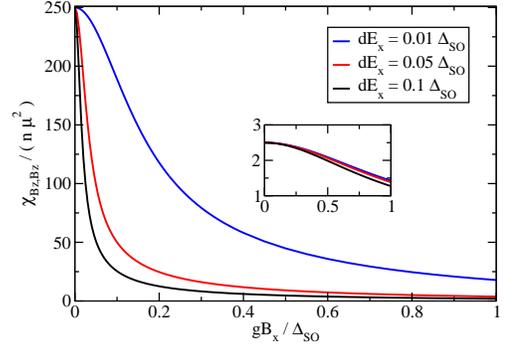}
\vspace{0.2cm}
\caption{
\label{fig:mszzbx} 
Magnetic susceptibility ($zz$ component) in  Eq. (\ref{eq:ms}) as a function of the magnetic
field in $x$ direction.  The three lines correspond to various values
of the external electric field in the $x$ direction.  The plot is for
the temperature $k\sub{B}T = 0.001\Delta\sub{SO}$.  The inset
represents the same quantity at a higher temperature $k\sub{B}T = 0.1\Delta\sub{SO}$.} 
\end{center}
\end{figure}

\begin{figure}[t]
\begin{center}
\includegraphics[width=6.5cm]{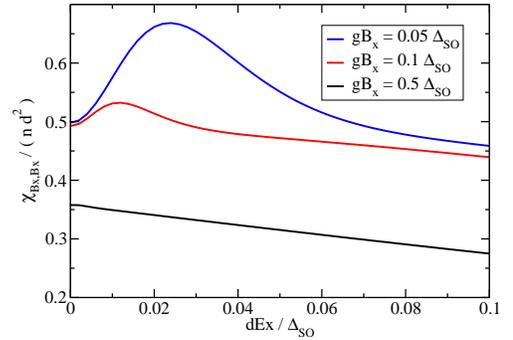}
\vspace{0.2cm}
\caption{\label{fig:msxxex}  Magnetic susceptibility ($xx$ component),
  Eq. (\ref{eq:ms}) as a function of the electric field in $x$ direction.  The three
  lines correspond to various values of the additional magnetic field in
  the $z$ direction.  The plot is for the temperature $k\sub{B}T =
  0.001\Delta\sub{SO}$.  } 
\end{center}
\end{figure}

The diagonal out-of-plane component of the magnetic susceptibility,
$\chi_{B_z,B_z}$, in the presence of an external magnetic field in the
$x$ direction decays strongly in the applied electric field along the
$x$ direction, Fig. \ref{fig:mszzbx}.  In electric fields, the
$\chi_{B_x,B_x}$ component shows a peak that is reduced by the
application of the magnetic field in $x$ direction,
Fig. \ref{fig:msxxex}.

We can derive the magnetic susceptibilities in the two regimes. In
the first case (i) we obtain (assuming now only linear effects in
$E$-field):
\begin{eqnarray}
\chi_{B_iB_j}&=&\frac{ng_{\perp}^2\mu_B^2}{2\Delta_{\mathcal{B}}}\bigg[\delta_{ij}-\frac{\mathcal{B}_i\mathcal{B}_j}{\Delta_{\mathcal{B}}^2}\nonumber\\ &+&\frac{\mathcal{E}\Delta\sub{SO}^2}{\mathcal{B}\Delta_{\mathcal{B}}^2}\left(\delta_{ij}-\frac{(3\mathcal{B}^2+\Delta_{\mathcal{B}}^2)\mathcal{B}_i\mathcal{B}_j}{B^2\Delta_{\mathcal{B}}^2}\right)\bigg],
\end{eqnarray}
with $i,j=x,y$, while
\begin{eqnarray}
\chi_{B_zB_z}&=&\frac{ng_z^2\mu_B^2}{2\Delta_{\mathcal
    B}}\frac{\Delta\sub{SO}^2}{\mathcal{B}\mathcal{E}},
\end{eqnarray}
for $\mathcal{B}_z\Delta\sub{SO}\ll \mathcal{B}\mathcal{E}$.  At low
temperatures the in-plane magnetic susceptibility shows a linear
dependence on the applied electric field $E$, thus allowing for a
simple estimate of the electric dipole parameter $d$ from magnetic
measurements.  Note that for strong electric fields
($\mathcal{E}\gg\mathcal{B}$), the magnetic susceptibility can vanish,
since the magnetization does not depend on the magnetic field
anymore. However, such a regime would not help to identify the
electric dipole coupling strength $d$ from susceptibility measurements
directly. The perpendicular magnetic susceptibility shows a strong
electric field dependence $\chi_{B_zB_z}\sim\mathcal{E}^{-1}$ and can
be used as an efficient probe for extracting  the electric dipole
parameter $d$. In the second case (ii) we obtain 
\begin{eqnarray}
\chi_{B_iB_j}&=&\frac{ng_{\perp}^2\mu_B^2}{2\Delta_{\mathcal{B}}}\bigg[\delta_{ij}-\frac{\mathcal{B}_i\mathcal{B}_j}{\Delta_{\mathcal{B}}^2}-\frac{\mathcal{E}^2\Delta\sub{SO}^2}{\Delta_{\mathcal{B}}^4}\nonumber\\ &\times&\left(\frac{3}{2}\left(\delta_{ij}+\frac{\mathcal{B}_i\mathcal{B}_j}{\Delta_{\mathcal{B}}^2}\right)-\beta\Delta_{\mathcal{B}}\left(\delta_{ij}+\frac{4\mathcal{B}_i\mathcal{B}_j}{\Delta_{\mathcal{B}}^2}\right)\right)\bigg],
\end{eqnarray}
when $i,j=x,y$, and 
\begin{eqnarray}
\chi_{B_zB_z}&=&\frac{ng_z^2\mu_B^2\mathcal{B}^2}{2\Delta_{\mathcal{B}}^3}\left(1+\beta\Delta_{\mathcal{B}}\frac{\Delta\sub{SO}^2}{\mathcal{B}^2}\right).
\end{eqnarray}
The magnetic response increases with temperature.  Also, in this
limit the dependence  of the magnetic susceptibility on the applied
electric field is rather weak
($\chi_{B_iB_j}(\mathcal{E})\sim\mathcal{E}^2$), thus this regime is
also not suitable for observing spin-electric effects.

For the magnetic susceptibility in the perpendicular magnetic field we find
\begin{equation}
\chi_{B_zB_z}=\frac{\beta ng_z^2\mu_B^2}{4}{\rm
  sech}(\beta\mathcal{B}),
\end{equation}
while for the in-plane magnetic susceptibility
$\chi_{B_{x(y)}B_{x(y)}}$ we obtain
\begin{eqnarray}
\chi_{B_{x(y)}B_{x(y)}}&=&\frac{ng_z^2\mu_B^2\Delta\sub{SO}}{2(\mathcal{B}^2-\Delta\sub{SO}^2)}\nonumber\\ &\times&\left[\frac{\mathcal{B}}{\Delta\sub{SO}}\left(1-\frac{\mathcal{E}^2}{\mathcal{B}^2}\right)\tanh{(\beta\mathcal{B})}-1\right],
\end{eqnarray}
in the limit $\mathcal{B},k_BT\ll \Delta\sub{SO}$. We mention that for
$B$ perpendicular to the molecular plane there is no electric field
$E$ (magnetic field $B$) dependence of the magnetization $M_i$
(electric polarization $P_i$). Thus, in order to see spin-electric
effects one needs to apply magnetic fields which have non-zero
in-plane components.

\subsection{Heat capacity}

Next we investigate the dependence of the heat capacity  on the
applied electric and magnetic  fields in different regimes. The heat
capacity is defined as $C=-\partial/\partial T(\partial
\ln{(Z)}/\partial \beta)$, so that we obtain
\begin{eqnarray}
C=\displaystyle \frac{Nk_B\beta^2}{4}\sum_{p=\pm
  1}\frac{(E_1+pE_{-1})^2}{\displaystyle\cosh^2{\left[\frac{\beta(E_1+pE_{-1})}{2}\right]}}.
\label{eq:hc}
\end{eqnarray}
We consider the cases of perpendicular B-field and in-plane B-field in
the limit $\Delta\sub{SO}\gg k_BT$. In the first case, i.e. for
$B\parallel z$ we obtain
\begin{eqnarray}
C\simeq Nk_B\beta^2\left\{\begin{array}{ll}
\displaystyle\Delta_{\mathcal{E}}^2e^{-2\beta\Delta\sub{SO}}+\mathcal{B}^2e^{-2\beta\mathcal{B}},
& \textrm{$\mathcal{B}\gg
  k_BT$}\vspace{0.2cm}\\ \displaystyle\Delta_{\mathcal{E}}^2e^{-2\beta\Delta\sub{SO}}+\frac{\mathcal{B}^2}{4},
& \textrm{$\mathcal{B}\ll k\sub{B}T$}.
\end{array}\right.
\end{eqnarray}  
The heat capacity $C$ shows a quadratic dependence on the applied
electric field for the entire range of E-field strengths. On the other
hand, the magnetic field dependence of $C$ is non-monotonic, and shows
a maximum for some finite B-field strength $\mathcal{B}_{max}\simeq
k_BT$.  In  the second situation, i.e. for $B\perp z$ we get
\begin{eqnarray}
C\simeq Nk_B\beta^2\left\{\begin{array}{ll} \displaystyle
{\frac{\mathcal{B}^2\mathcal{E}^2}{\Delta_{\mathcal{B}}^2}e^{-2\frac{\beta\mathcal{B}\mathcal{E}}{\Delta_{\mathcal{B}}}}+\Delta_{\mathcal{B}}^2e^{-2\beta\Delta_{\mathcal{B}}}},
&  \textrm{$\mathcal{E}\gg k_BT$}\vspace{0.2cm}\\ \displaystyle
\frac{\mathcal{B}^2\mathcal{E}^2}{4\Delta_{\mathcal{B}}^2}, &
\textrm{$\mathcal{E}\ll k_BT$}.
\end{array}\right.
\end{eqnarray}  
As in the previous case, the dependence of the heat-capacity $C$ is
linear in E-field for low E-fields. However, for large E-fields the
dependence is non-monotonic and thus shows a maximum for some finite
electric field strength $\mathcal{E}_{max}\simeq k_BT$. Note that in
this case also the dependence on the magnetic field is non-monotonic,
and thus we obtain a second maximum for $\mathcal{B}_{max}\simeq
k_BT$. We can conclude from the above expressions that the strongest
dependence of the heat capacity on the electric field is when the
magnetic field is applied in-plane, and then it is mostly quadratic.

\begin{figure}[t]
\begin{center}
\includegraphics[width=6.5cm]{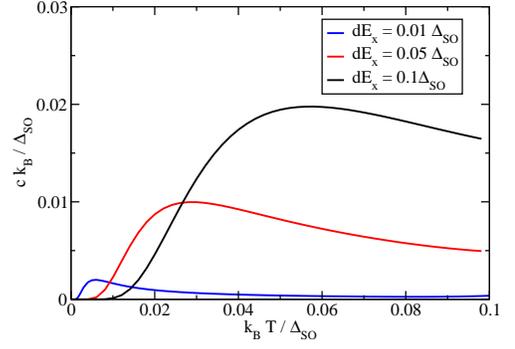}
\vspace{0.2cm}
\caption{\label{fig:ce}  Heat capacity, Eq. (\ref{eq:hc}), as a function of
  temperature in various electric fields.  } 
\end{center}
\end{figure}  
\begin{figure}[t]
\begin{center} 
\includegraphics[width=6.5cm]{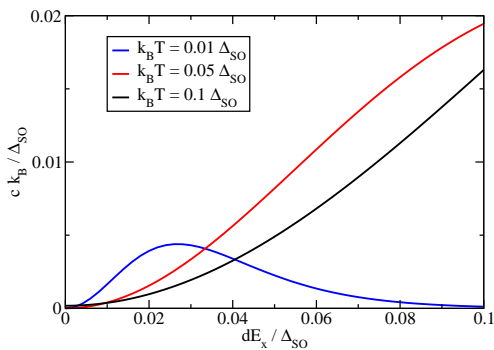}
\vspace{0.2cm}
\caption{\label{fig:esyyby}  Heat capacity, Eq. (\ref{eq:hc}),  at low temperature as a
  function of external electric field.} 
\end{center}
\end{figure}  

For the derivation of all the thermodynamic quantities presented in
the previous sections, we have restricted ourselves to the
contributions arising from only the lowest four states, even though
the spin system spans eight states in total. This description is valid
if the splitting between the energies of $S=1/2$ and $S=3/2$ states is
much larger than the temperature $k\sub{B}T$.  This splitting varies
strongly with the applied magnetic field, for $\mathcal{B}=3J/4$ one
of the $S=3/2$ states ($M=-3/2$) crossing the $M=1/2$ of the $S=1/2$
states and, even more, for $\mathcal{B}>3J/2$ the $M=-3/2$ becomes the
spin system ground states. Thus, for large magnetic fields our
effective description in terms of only the $S=1/2$ states breaks down
and one have to reconsider the previous quantities in this limit.

\section{Conclusions}

Electric fields can be applied at very short spatial and temporal
scales which makes them preferable  for quantum information processing
applications  over the more  standard magnetic fields .  Nanoscale
magnets, while displaying  rich quantum dynamics, have not yet been
shown to respond to electric fields in experiments.  We have
investigated theoretically  the possibility of spin-electric coupling
in nanoscale magnets using symmetry analysis, and found that the
spin-electric coupling is possible in antiferromagnetic ground-state
manifolds of spin-$1/2$ and spin-$3/2$ triangles, as well as in
spin-$1/2$ pentagon.  The spin-electric coupling in the triangle can
exist even in the absence of spin-orbit coupling, while the coupling
in the pentagon requires the spin-orbit interaction in the molecule.
We have characterized the form of the spin-electric coupling in all of these
molecules and presented the selection rules for the transitions
between the spin states induced by electric fields.  

While the symmetry can predict the presence or absence of the
spin-electric coupling, it can not predict the size of the
corresponding coupling constant.  In order to find a molecule suitable
for electric manipulation, it is necessary to have an estimate of the
spin-electric coupling strength.  For this purpose, we have described
the nanoscale magnets in terms of the Hubbard model, and related the
coupling constants of the symmetry-based models to the hopping and
on-site energy parameters of the Hubbard model.  We have found that
the modification of the Hubbard model parameters due to the electric
field produces a spin-electric coupling of the same form as predicted
by the symmetry analysis.  However, within the Hubbard model, the
coupling constants have a clear and intuitive meaning in terms of the
hopping and on-site energies of the localized electrons.  We have also
studied the superexchange interaction of the spins on the magnetic
centers through the bridge.  If we assume that the interaction of the
localized spins is a property of the bridge alone, the spin-electric
coupling can be calculated by ab-initio analysis of the bridge alone,
and not of the entire molecule.

Finally, we analyzed the role of spin-electric coupling in standard
experimental setups typically used for the characterization of
nanoscale magnets.  We find that the spin-electric coupling can be
detected in the ESR and NMR spectra that probe the local spins.  Also,
thermodynamic quantities, like the polarization, magnetization, linear
magnetoelectric effect, and the specific heat show signatures of
spin-electric coupling in the triangular molecules.  Thus, our results
set a path toward finding  suitable molecules that exhibit
spin-electric effects and how they can be identified experimentally.

In this work, we have focused on the spin rings with an odd number of
magnetic centers (odd spin rings), whose low-energy spectrum is
dominated by frustration effects.  The odd spin rings, due to
frustration, possess a four-fold degenerate ground state multiplet,
which can be split by electric fields.  As opposed to the odd spin
rings, the ground states of even-spin rings is usually a
non-degenerate $S=0$ state, separated from the higher energy states by
a gap of the order of exchange coupling $J$.  Coupling of the electric
field to these states can thus proceed only via excited states, and
the coupling strength is reduced by $d|\vctr{E}|/J$.  Similarly, in
lower-symmetry odd-spin rings, the ground state multiplet consists of
an $S=1/2$ Kramers doublet, which can not be split by electric fields,
i.e. there is no spin-electric effect in zero magnetic field.
Therefore, the odd spin rings seem to be the most suitable candidates
for observing the spin-electric coupling and using it to control the
spins.

We thank M. Affronte and V. Bellini for useful discussions.  We
acknowledge financial support from the Swiss NSF, the NCCR Nanoscience
Basel, the Italian MIUR under FIRB Contract No. RBIN01EY74, and the EU
under "MagMaNet" and "MolSpinQIP".

\appendix

\section{Spin states in terms of the $c^{\dagger}_{\Gamma}$ operators}

In this appendix we show the expressions for the three-electron symmetry adapted states $|\psi_{\Gamma}^{i,\sigma}\rangle$ in Eqs. (\ref{singly_occupied}) and (\ref{doubly_occupied}) in terms of the  symmetry adapted creation operators $c^{\dagger}_{\Gamma,\sigma}$. Making use of Eq. (\ref{symmetry_operators}) we obtain
\begin{eqnarray}
|\psi_{A_2^{'}}^{1\sigma}\rangle&=&\frac{i\epsilon}{\sqrt{3}}\big(c^{\dagger}_{A_1^{'}\bar{\sigma}}c^{\dagger}_{E_+^{'}\sigma}c^{\dagger}_{E_-^{'}\sigma}+c^{\dagger}_{E_+^{'}\bar{\sigma}}c^{\dagger}_{E_-^{'}\sigma}c^{\dagger}_{A_1^{'}\sigma}\nonumber\\&-&c^{\dagger}_{E_-^{'}\bar{\sigma}}c^{\dagger}_{E_+^{'}\sigma}c^{\dagger}_{A_1^{'}\sigma}\big)|0\rangle\\ 
|\psi_{E_+^{'}}^{1\sigma}\rangle&=&\frac{i}{\sqrt{3}}\big(c^{\dagger}_{A_1^{'}\bar{\sigma}}c^{\dagger}_{A_1^{'}\sigma}c^{\dagger}_{E_+^{'}\sigma}+\epsilon
c^{\dagger}_{E_+^{'}\bar{\sigma}}c^{\dagger}_{E_+^{'}\sigma}c^{\dagger}_{E_-^{'}\sigma}\nonumber\\&+&\bar{\epsilon}c^{\dagger}_{E_-^{'}\bar{\sigma}}c^{\dagger}_{E_-^{'}\sigma}c^{\dagger}_{A_1^{'}\sigma}\big)|0\rangle\\ 
|\psi_{E_-^{'}}^{1\sigma}\rangle&=&\frac{i}{\sqrt{3}}\big(c^{\dagger}_{A_1^{'}\bar{\sigma}}c^{\dagger}_{A_1^{'}\sigma}c^{\dagger}_{E_-^{'}\sigma}+\epsilon
c^{\dagger}_{E_-^{'}\bar{\sigma}}c^{\dagger}_{E_-^{'}\sigma}c^{\dagger}_{E_+^{'}\sigma}\nonumber\\&+&\bar{\epsilon}c^{\dagger}_{E_+^{'}\bar{\sigma}}c^{\dagger}_{E_+^{'}\sigma}c^{\dagger}_{A_1^{'}\sigma}\big)|0\rangle\\ 
|\psi_{A_1^{'}}^{2\sigma}\rangle&=&\frac{\sigma\epsilon}{\sqrt{2}}\left(c^{\dagger}_{E_+^{'}\bar{\sigma}}c^{\dagger}_{A_1^{'}\sigma}c^{\dagger}_{E_-^{'}\sigma}+c^{\dagger}_{E_-^{'}\bar{\sigma}}c^{\dagger}_{A_1^{'}\sigma}c^{\dagger}_{E_+^{'}\sigma}\right)|0\rangle\\ 
|\psi_{A_2^{'}}^{2\sigma}\rangle&=&-\frac{i\sigma\epsilon}{\sqrt{6}}\big(2c^{\dagger}_{A_1^{'}\bar{\sigma}}c^{\dagger}_{E_+^{'}\sigma}c^{\dagger}_{E_-^{'}\sigma}+c^{\dagger}_{E_+^{'}\bar{\sigma}}c^{\dagger}_{A_1^{'}\sigma}c^{\dagger}_{E_-^{'}\sigma}\nonumber\\&-&c^{\dagger}_{E_-^{'}\bar{\sigma}}c^{\dagger}_{A_1^{'}\sigma}c^{\dagger}_{E_+^{'}\sigma}\big)|0\rangle\\ 
|\psi_{E^{'1}_{+}}^{2\sigma}\rangle&=&\frac{\sigma}{\sqrt{2}}\left(\bar{\epsilon}c^{\dagger}_{A_1^{'}\bar{\sigma}}c^{\dagger}_{A_1^{'}\sigma}c^{\dagger}_{E_+^{'}\sigma}+\epsilon
c^{\dagger}_{E_-^{'}\bar{\sigma}}c^{\dagger}_{A_1^{'}\sigma}c^{\dagger}_{E_-^{'}\sigma}\right)|0\rangle\\ 
|\psi_{E^{'1}_{-}}^{2\sigma}\rangle&=&\frac{\sigma}{\sqrt{2}}\left(\bar{\epsilon}c^{\dagger}_{A_1^{'}\bar{\sigma}}c^{\dagger}_{A_1^{'}\sigma}c^{\dagger}_{E_-^{'}\sigma}+\epsilon
c^{\dagger}_{E_+^{'}\bar{\sigma}}c^{\dagger}_{A_1^{'}\sigma}c^{\dagger}_{E_+^{'}\sigma}\right)|0\rangle\\ 
|\psi_{E^{'2}_{+}}^{2\sigma}\rangle&=&\frac{i\sigma\bar{\epsilon}}{\sqrt{6}}\big(c^{\dagger}_{A_1^{'}\bar{\sigma}}c^{\dagger}_{A_1^{'}\sigma}c^{\dagger}_{E_+^{'}\sigma}-\bar{\epsilon}c^{\dagger}_{E_-^{'}\bar{\sigma}}c^{\dagger}_{A_1^{'}\sigma}c^{\dagger}_{E_-^{'}\sigma}\nonumber\\&-&2\epsilon
c^{\dagger}_{E_+^{'}\bar{\sigma}}c^{\dagger}_{E_+^{'}\sigma}c^{\dagger}_{E_-^{'}\sigma}\big)|0\rangle\\ 
|\psi_{E^{'2}_{-}}^{2\sigma}\rangle&=&\frac{i\sigma\bar{\epsilon}}{\sqrt{6}}\big(c^{\dagger}_{A_1^{'}\bar{\sigma}}c^{\dagger}_{A_1^{'}\sigma}c^{\dagger}_{E_-^{'}\sigma}-\bar{\epsilon}c^{\dagger}_{E_+^{'}\bar{\sigma}}c^{\dagger}_{A_1^{'}\sigma}c^{\dagger}_{E_+^{'}\sigma}\nonumber\\&-&2\epsilon
c^{\dagger}_{E_-^{'}\bar{\sigma}}c^{\dagger}_{E_-^{'}\sigma}c^{\dagger}_{E_+^{'}\sigma}\big)|0\rangle,
\end{eqnarray}
where $\sigma$ stands above for $sign(\sigma)$.

\section{$H\sub{SO}$, $H\sub{e-d}^0$, and $H\sub{e-d}^1$ matrix elements}

For the SOI matrix elements we obtain
\begin{eqnarray}
\langle\psi_{A_1^{'}}^{2\sigma}|H\sub{SO}|\psi_{A_2^{'}}^{1\sigma}\rangle&=&\frac{2i\lambda\sub{SO}}{\sqrt{2}}\sigma,\\ \langle\psi_{E^{'1}_{\pm}}^{2\sigma}|H\sub{SO}|\psi_{E_{\pm}^{'}}^{1\sigma}\rangle&=&\pm\frac{i\bar{\epsilon}\lambda\sub{SO}}{\sqrt{2}}\sigma,\\ \langle\psi_{E^{'2}_{\pm}}^{2\sigma}|H\sub{SO}|\psi_{E_{\pm}^{'}}^{1\sigma}\rangle&=&\pm\sigma\frac{\sqrt{3}\epsilon\lambda\sub{SO}}{\sqrt{2}}\sigma,\\ \langle\psi_{A_1^{'}}^{2\sigma}|H\sub{SO}|\psi_{A_2^{'}}^{2\sigma}\rangle&=&-\sigma2\lambda\sub{SO}\\ \langle\psi_{E^{'1}_{\pm}}^{2\sigma}|H\sub{SO}|\psi_{E^{'1}_{\pm}}^{2\sigma}\rangle&=&\pm\sigma\frac{\sqrt{3}}{2}\lambda\sub{SO},\\ \langle\psi_{E^{'1}_{\pm}}^{2\sigma}|H\sub{SO}|\psi_{E^{'2}_{\pm}}^{2\sigma}\rangle&=&\pm\frac{i\lambda\sub{SO}}{2}\sigma,\\ \langle\psi_{E^{'2}_{\pm}}^{2\sigma}|H\sub{SO}|\psi_{E^{'2}_{\pm}}^{2\sigma}\rangle&=&\mp\sigma\frac{\sqrt{3}}{2}\lambda\sub{SO},
\end{eqnarray}
while  the remaining terms are equal to zero. For the electric dipole matrix
elements we obtain
\begin{eqnarray}
\langle\psi_{E^{'1}_-}^{2\sigma}|H\sub{e-d}^0|\psi_{E^{'1}_+}^{2\sigma}\rangle&=&\frac{a}{2}\left((\bar{\epsilon}-1)E_x+\epsilon\sqrt{3}
E_y\right)\\ 
\langle\psi_{E^{'2}_-}^{2\sigma}|H\sub{e-d}^0|\psi_{E^{'2}_+}^{2\sigma}\rangle&=&\frac{a}{2}\left(\epsilon
E_x+\frac{1-\bar{\epsilon}}{\sqrt{3}}
E_y\right)\\ 
\langle\psi_{E^{'1}_-}^{2\sigma}|H\sub{e-d}^0|\psi_{E^{'2}_+}^{2\sigma}\rangle&=&-\frac{a}{2}\left(\epsilon
E_x+\frac{1-\bar{\epsilon}}{\sqrt{3}}E_y\right)\\ 
\langle\psi_{E^{'2}_-}^{2\sigma}|H\sub{e-d}^0|\psi_{E^{'1}_+}^{2\sigma}\rangle&=&-\frac{a}{2}\left((\bar{\epsilon}-1)E_x+\epsilon\sqrt{3}
E_y\right)\\ 
\langle\psi_{E_-^{'}}^{1\sigma}|H\sub{e-d}^1|\psi_{E^{'1}_+}^{2\sigma}\rangle&=&-\frac{i\epsilon
  E}{\sqrt{6}}(\epsilon
d^*_{EE}-2\bar{\epsilon}d^*_{AE}-d_{AE})\\\langle\psi_{E_+^{'}}^{1\sigma}|H\sub{e-d}^1|\psi_{E^{'1}_-}^{2\sigma}\rangle&=&\frac{i\epsilon
  \bar{E}}{\sqrt{6}}(\epsilon d_{EE}+2\bar{\epsilon}
d^*_{AE}+d_{AE}),\\ \langle\psi_{E_-^{'}}^{1\sigma}|H\sub{e-d}^1|\psi_{E^{'2}_+}^{2\sigma}\rangle&=&\frac{\epsilon
  E}{\sqrt{2}}(\epsilon
d^*_{EE}+d_{AE}),\\ \langle\psi_{E_+^{'}}^{1\sigma}|H\sub{e-d}^1|\psi_{E^{'2}_-}^{2\sigma}\rangle&=&-\frac{\epsilon
  \bar{E}}{\sqrt{2}}(\epsilon d_{EE}-
d_{AE})\\ \langle\psi_{E_-^{'}}^{1\sigma}|H\sub{e-d}^1|\psi_{E^{'1}_+}^{1\sigma}\rangle&=&0.
\end{eqnarray}


\end{document}